\documentclass[11pt,a4paper]{article}
\pdfoutput=1

\usepackage{jheppub}

\usepackage{amsmath}
\usepackage{bbm}
\usepackage{float}
\usepackage{graphicx}	
\usepackage{hyperref}
\usepackage{mathtools}
\usepackage{cancel}

\usepackage[normalem]{ulem}

\usepackage{multirow}
\usepackage{rotating}
\usepackage{subcaption}

\def\lsim{\raise0.3ex\hbox{$\;<$\kern-0.75em\raise-1.1ex
\hbox{$\sim\;$}}}
\def\gsim{\raise0.3ex\hbox{$\;>$\kern-0.75em\raise-1.1ex
\hbox{$\sim\;$}}}

\def\thetitle{ 
Neutrino amplitude decomposition: Toward observing the atmospheric - solar wave interference \\
 \vspace{- 6mm}
}
\title{\thetitle}
\hypersetup{pdftitle={\thetitle}}
\author{Hisakazu Minakata}
\affiliation{
Center for Neutrino Physics, Department of Physics, Virginia Tech, Blacksburg, Virginia 24061, USA \\
}

\emailAdd{minakata71@vt.edu}
\date{\today}

\abstract{ 
Observation of the interference between the atmospheric and solar oscillation waves with the correct magnitude would provide another manifestation of the three-generation structure of leptons. As a prerequisite for such analyses we develop a method for decomposing the oscillation $S$ matrix into the atmospheric and solar amplitudes. Though the similar method was recently proposed successfully in vacuum, once an extension into the matter environment is attempted, it poses highly nontrivial problems. Even for an infinitesimal matter potential, inherent mixture of the atmospheric and solar oscillation waves occurs, rendering a simple extension of the vacuum definition untenable. We utilize general kinematic structure as well as analyses of the five perturbative frameworks, in which the nature of matter-dressed atmospheric and solar oscillations are known, to understand the origin of the trouble, how to deal with the difficulty, and to grasp the principle of decomposition. Then, we derive the amplitude decomposition formulas in these frameworks, and discuss properties of the decomposed probabilities. We mostly discuss the $\nu_{\mu} \rightarrow \nu_{e}$ channel, but a comparison with the $\nu_{\mu} \rightarrow \nu_{\tau}$ channel reveals an interesting difference. 
}

\begin{document} % JHEP 

\maketitle

\section{Introduction}
\label{sec:introduction} 

The three-generation structure of the fundamental fermions, leptons and quarks, is one of the most salient features in our world. Most notably, it has a dramatic consequence that CP symmetry must be broken \cite{Kobayashi:1973fv}, barring exceptional values of the CP phase. CP violation was indeed observed experimentally \cite{Christenson:1964fg}, and its origin \'a la Kobayashi-Maskawa mechanism was confirmed \cite{Aubert:2001nu,Abe:2001xe}. It is strongly suspected that the similar structure is endowed also in the lepton sector, and by now there exists an evidence for CP violation at a confidence level (CL) close to $3\sigma$ \cite{Abe:2019vii}.

In a previous paper \cite{Huber:2019frh}, we have argued that as one of the other consequences of the three-flavor structure, a highly nontrivial one, we must be able to observe quantum interference between the atmospheric and the solar oscillation amplitudes. We think it a very interesting point, bridging between the remarkable structure of the fundamental fermions and the quantum mechanical nature of the phenomenon of neutrino oscillation, which is subsumed into the most successful quantum field theory, the neutrino-mass-embedded Standard Model ($\nu$SM). 

To illuminate the point, we took a concrete setting of the medium-baseline reactor neutrino experiment JUNO \cite{An:2015jdp} to simulate the data set and demonstrated, with careful implementation of the systematic errors, that it will be able to detect the interference effect between the atmospheric and solar amplitudes at a CL higher than $4\sigma$~\cite{Huber:2019frh}. Since $\bar{\nu}_{e}$ (or $\nu_e$) disappearance channel is free from the CP phase $\delta$ in vacuum as well as in matter \cite{Kuo:1987km,Minakata:1999ze}, the nature of interference which will be observed by JUNO is completely free from effects of the phase $\delta$, in sharp contrast to the situation expected in the accelerator neutrino appearance measurements.

Despite its fundamental importance, to our knowledge, this topic did not appear to receive a sufficient attention in our community to a level it deserves. See, however, refs.~\cite{Smirnov:2006sm,Akhmedov:2008qt,Nunokawa:2007qh,Petcov:2001sy,Choubey:2003qx,Learned:2006wy} for the relatively few foregoing works. 
It is certainly possible that the shortage of the list simply reflects our ignorance. But it is difficult to make a complete list of the foregoing works that addressed the interference effect due to a too broad spectrum which spans from its implicit treatment to the discussion of the isolated effect of interference.\footnote{ 
%%%%%%%%%%%%% footnote %%%%%%%%%%%%%% 
Every analysis of neutrino data with accurate integration of neutrino propagation equation automatically contains the interference effect between the atmospheric and solar oscillation waves. Likewise, discussion of sub-leading $\Delta m^2_{21}$ effect in regions with the dominant $\Delta m^2_{31}$ effect \cite{Fogli:1998au,Peres:1999yi,Peres:2003wd}, %% sub-leading-21
or vise versa \cite{Gando:2010aa,Abe:2016nxk},   %% sub-leading-31, 
inevitably contain the interference effect. On the other hand, we are talking about how to extract the interference term in the probability and the physical properties of the isolated interference term in this and the previous papers. 
}

We have prescribed in ref.~\cite{Huber:2019frh} the way of how the oscillation $S$ matrix can be decomposed into the atmospheric and the solar amplitudes in vacuum, the indispensable first step to discuss the interference effect. Hereafter, we refer this procedure as the {\em ``amplitude decomposition''}, the terminology which will also be used for the case in matter. The definition of the decomposed amplitudes includes the completeness condition $S_{\alpha \beta} = \delta_{\alpha \beta} + S_{\alpha \beta}^{ \text{atm} } + S_{\alpha \beta}^{ \text{sol} }$, with $\alpha, \beta$ being the flavor indices. This is nothing but a manifestation of the three generation structure of neutrinos, namely, presence of only the two independent modes of oscillation with the two different frequencies. 
Notice that though it is often stated that the three-flavor structure of neutrinos is well known, in fact, it is not known {\em at all} whether it is sufficient or not. Therefore, observing the interference term with the correct magnitude dictated by the $\nu$SM provides a new form of unitarity test. 

In this paper, we discuss the amplitude decomposition in matter. In most of the neutrino experiments, done with use of the accelerator, atmospheric, solar, or even the reactor neutrinos, neutrinos pass through matter, thereby receiving the matter effect \cite{Wolfenstein:1977ue,Mikheev:1986gs}. Though the effect may be small for the low-energy reactor neutrinos, it is comparable to the vacuum oscillation effect, for example, in the ongoing and upcoming long-baseline (LBL) accelerator neutrino experiments, T2K~\cite{Abe:2019vii}, MINOS/MINOS+~\cite{Adamson:2020ypy}, NO$\nu$A~\cite{Acero:2019ksn}, T2HK~\cite{Abe:2018uyc}, and DUNE~\cite{Abi:2020evt}. To detect the interference effects between the atmospheric and solar oscillation waves in a quantitative manner, and to understand physics involved in it, we have to isolate the interference term in the probability first. Hence, the amplitude decomposition is an indispensable machinery in our approach. 

Upon turning on the matter potential, however, albeit with an infinitesimal magnitude, we immediately encounter a difficulty. What happens is that the matter effect mixes the $\Delta m^2_{31}$- and $\Delta m^2_{21}$-driven waves, and this genuine three-flavor effect makes a simple extension of the vacuum definition of the amplitude decomposition untenable. Since the energy eigenvalues are unaffected with the infinitesimal matter potential, it represents the inherent difficulty of amplitude decomposition in matter. Thus, we face, from the beginning, with the conceptual difficulty in extending our vacuum definition of the amplitude decomposition into that in matter.

Since the ``atmospheric'' and the ``solar'' waves are generally modified by the effects of the matter potential, identification and separation of these two modes are highly nontrivial problem in matter. In this paper, therefore, we first try to find obstacles to perform the amplitude decomposition in matter, understand the problems, and solve them if possible. We then analyze several perturbative frameworks in which nature of the matter-modified atmospheric and solar oscillations are reasonably understood to learn how we reach the prescription for the amplitude decomposition. In section~\ref{sec:approach}, we explain more about how we approach these conceptually involved and technically non-tractable problem of amplitude decomposition in matter.

In this paper, we will have to take a several different routes to proceed toward the end with many corners to turn. The organization of this paper is better explained in some of the corners at which we make a turn. Presentation in this paper will be very pedagogical, as it may be appropriate for the subject for which no systematic treatment is available to our knowledge. We leave most of the technical discussions to appendices. An essence of this paper, or at least what we try to achieve in this paper, can be grasped in reading section~\ref{sec:approach} of only one page.

\section{Amplitude decomposition in vacuum} 
\label{sec:decomposition-vacuum}

We start by discussing decomposition of the $S$ matrix into the atmospheric and the solar amplitudes in vacuum, mostly recollecting what we have done in \cite{Huber:2019frh}. For simplicity, we introduce the compact notations for the oscillation phase variables 
\begin{eqnarray} 
\Delta m^2_{ji} \equiv m_{j}^2 - m_{i}^2, 
\hspace{10mm}
\Delta_{ji} \equiv \frac{ \Delta m^2_{ji} }{ 2E }  
\hspace{6mm} 
(i, j = 1,2,3), 
\label{Delta-ij-def}
\end{eqnarray}
where $m_{i}$ denotes the mass of the $i$-th eigenstate neutrino and $E$ is the neutrino energy. The notations will be used throughout this paper.

\subsection{Amplitude decomposition: Heuristic method} 
\label{sec:heuristic-method} 

The neutrino oscillation $S$ matrix element which describes the neutrino oscillation $\nu_\beta \rightarrow \nu_\alpha$ ($\alpha \neq \beta$, or $\alpha = \beta$) in vacuum, 
\begin{eqnarray} 
S_{\alpha \beta} = 
U_{\alpha 1} U^{*}_{\beta 1} e^{ - i \frac{ m^2_{1} }{ 2E } x }
+ U_{\alpha 2} U^{*}_{\beta 2} e^{ - i \frac{ m^2_{2} }{ 2E } x }
+ U_{\alpha 3} U^{*}_{\beta 3} e^{ - i \frac{ m^2_{3} }{ 2E } x }, 
\label{S-matrix-def-original}
\end{eqnarray}
can be written, after redefining the phase by removing $e^{ - i ( m_{1}^2 / 2E ) x}$ as
\begin{eqnarray} 
S_{\alpha \beta} = 
U_{\alpha 1} U^{*}_{\beta 1}  
+ U_{\alpha 2} U^{*}_{\beta 2} e^{ - i \Delta_{21} x} 
+ U_{\alpha 3} U^{*}_{\beta 3} e^{ - i \Delta_{31} x}, 
\label{S-matrix-def}
\end{eqnarray}
where $U \equiv U_{\text{\tiny MNS}}$ denotes the lepton flavor mixing matrix~\cite{Maki:1962mu}. We use, apart from section~\ref{sec:DMP}, the Particle Data Group (PDG) convention of $U_{\text{\tiny MNS}}$ \cite{Tanabashi:2018oca}, see eq.~\eqref{MNSmatrix-PDG}. By using unitarity of the $U$ matrix, $\sum_{i} U_{\alpha i} U^{*}_{\beta i} = \delta_{\alpha \beta}$, $S_{\alpha \beta}$ can be written as \cite{Nunokawa:2007qh,Bilenky:2012zp,Huber:2019frh}
\begin{eqnarray} 
S_{\alpha \beta} = 
\delta_{\alpha \beta} 
+ U_{\alpha 2} U^{*}_{\beta 2} 
\left( e^{ - i \Delta_{21} x} - 1 \right) 
+ U_{\alpha 3} U^{*}_{\beta 3} 
\left( e^{ - i \Delta_{31} x} - 1 \right) 
\label{S-matrix-decompose}
\end{eqnarray}
where $\delta_{\alpha \beta}$ denotes the Kronecker delta function. Equation~\eqref{S-matrix-decompose} defines the atmospheric and the solar amplitudes 
\begin{eqnarray} 
&& 
S_{\alpha \beta}^{ \text{atm} } 
\equiv  
U_{\alpha 3} U^{*}_{\beta 3} \left( e^{ - i \Delta_{31} x} - 1 \right), 
\nonumber \\
&& 
S_{\alpha \beta}^{ \text{sol} } \equiv  
U_{\alpha 2} U^{*}_{\beta 2}  \left( e^{ - i \Delta_{21} x} -1 \right).
\label{atm-sol-amplitude-vac}
\end{eqnarray}
Though the above procedure might look ad hoc, one can define the atmospheric and the solar amplitudes in a more systematic way. 

\subsection{Definition of the decomposed amplitudes in vacuum} 
\label{sec:amplitude-decomposition-vacuum} 

Let us give the general definition 1 of amplitude decomposition in vacuum and require the completeness condition 2. In fact, we even try to apply the same definition in an environment in matter, when we talk about the decomposition in the narrow sense, where the atmospheric and solar wave are defined to be $\Delta m^2_{31}$- and $\Delta m^2_{21}$-driven oscillations, respectively. 
\begin{enumerate} 
\item
For a given $S$ matrix element $S_{\alpha \beta}$, the atmospheric and the solar amplitudes are defined, respectively, as\footnote{
%%%%%%%%%%%%% footnote %%%%%%%%%%%%%%
The limit used in \eqref{atm-sol-amplitude-vac-def} is to define the amplitude decomposition, not the statement that $\Delta m^2_{21}$ is approximately small. In vacuum the definition applies even in the case $\vert \Delta m^2_{31} \vert < \Delta m^2_{21}$. 
}
\begin{eqnarray} 
S_{\alpha \beta}^{ \text{atm} } 
= \lim_{\Delta_{21} \rightarrow 0} S_{\alpha \beta} - \delta_{\alpha \beta}, 
\hspace{10mm}
S_{\alpha \beta}^{ \text{sol} } 
= \lim_{\Delta_{31} \rightarrow 0} S_{\alpha \beta} - \delta_{\alpha \beta}.
\label{atm-sol-amplitude-vac-def} 
\end{eqnarray}
A consistency check on the obtained amplitudes is that they must satisfy 
\begin{eqnarray} 
\lim_{\Delta_{31} \rightarrow 0} S_{\alpha \beta}^{ \text{atm} } 
= \lim_{\Delta_{21} \rightarrow 0} S_{\alpha \beta}^{ \text{sol} } = 0.
\label{consistency}
\end{eqnarray}

\item 
We demand the completeness condition 
\begin{eqnarray} 
S_{\alpha \beta} = \delta_{\alpha \beta} + S_{\alpha \beta}^{ \text{atm} } + S_{\alpha \beta}^{ \text{sol} }.
\label{completeness}
\end{eqnarray}

\end{enumerate}
We have shown in vacuum that the procedure reproduces the decomposition in~\eqref{atm-sol-amplitude-vac} \cite{Huber:2019frh}. 

The general definition 1 of the atmospheric and the solar amplitudes and the completeness condition 2 are natural to require. The atmospheric amplitude, by definition, describes neutrino oscillation due to non-vanishing $\Delta m^2_{31}$, and the solar amplitude the one caused by $\Delta m^2_{21}$. The definition 1 just reflects this feature together with the consistency condition that $S_{\alpha \beta}^{ \text{atm} }$ ($S_{\alpha \beta}^{ \text{sol} }$) must vanish if $\Delta m^2_{31}=0$ ($\Delta m^2_{21}=0$).
The condition 2 requires that decomposition of the oscillation amplitude into the atmospheric and the solar amplitudes must be complete. It reflects the fact that only the two independent $\Delta m^2$ are available, the atmospheric $\Delta m^2_{31}$ and the solar $\Delta m^2_{21}$, and hence only the two independent amplitudes exist, a manifestation of the three generation mixing.

\subsection{Interference terms in the probability: Comparison between the $\nu_\mu - \nu_e$ channel and the ones in the $\nu_\mu - \nu_\tau$ sector} 
\label{sec:mue-vs-mutau} 

We have discussed in ref.~\cite{Huber:2019frh} the amplitude decomposition in vacuum, and exhibited the explicit forms of the non-interference and interference terms in the probability in the $\nu_{e}$ related sector. Here we present the similar results in the $\nu_\mu - \nu_\tau$ sector, and compare them to the one in the $\nu_\mu - \nu_e$ channel. Our focus is mainly on the appearance channels. It will reveal a new feature of the ingredient in the interference term. For convenience of our discussion, we partly recapitulate the features of the probability in the $\nu_\mu - \nu_e$ channel. 

In the $\nu_{\mu} \rightarrow \nu_{e}$ channel the decomposed atmospheric and solar amplitudes read 
\begin{eqnarray} 
S^{ \text{atm} }_{e \mu} 
&=&
s_{23} e^{ - i \delta} c_{13} s_{13} 
\left( e^{ - i \Delta_{31} x } - 1 \right), 
\nonumber \\ 
S^{ \text{sol} }_{e \mu} 
&=& 
c_{13} s_{12} 
\left( c_{23} c_{12} - s_{23} s_{12} s_{13} e^{ - i \delta} \right) 
\left( e^{ - i \Delta_{21} x } - 1 \right).
\label{Semu-atm-sol-vacuum} 
\end{eqnarray}
The amplitude decomposition \eqref{atm-sol-amplitude-vac} leads to the decomposed probability 
\begin{eqnarray} 
&&
P(\nu_{\beta} \rightarrow \nu_{\alpha}) 
= P(\nu_{\beta} \rightarrow \nu_{\alpha})^{ \text{non-int-fer} }
+ P(\nu_{\beta} \rightarrow \nu_{\alpha})^{ \text{int-fer} }.
\label{probability-beta-alpha} 
\end{eqnarray}
With \eqref{Semu-atm-sol-vacuum}, the non-interference and interference parts of the probability in the $\nu_{\mu} \rightarrow \nu_{e}$ channel are given by \cite{Huber:2019frh}
\begin{eqnarray} 
&& P(\nu_{\mu} \rightarrow \nu_{e})^{ \text{non-int-fer} } 
\equiv 
\vert S_{e \mu}^{ \text{atm} } \vert^2 
+ \vert S_{e \mu}^{ \text{sol} } \vert^2 
= 
s^2_{23} \sin^2 2 \theta_{13} \sin^2 \frac{ \Delta_{31} x }{2}
\nonumber \\ 
&+&
\biggl[ 
c^2_{23} c^2_{13} \sin^2 2\theta_{12} 
+ s^2_{23} s^4_{12} \sin^2 2\theta_{13} 
- 8 s^2_{12} J_r \cos \delta 
\biggr] 
\sin^2 \frac{ \Delta_{21} x }{2}, 
\nonumber \\
&&
P(\nu_{\mu} \rightarrow \nu_{e})^{ \text{int-fer} } 
\equiv 
2 \mbox{Re} \left[ \left( S_{e \mu}^{ \text{atm} } \right)^* S_{e \mu}^{ \text{sol} } \right] 
\nonumber \\ 
&=&
8 \biggl[
\left( J_r \cos \delta - s^2_{23} c^2_{13} s^2_{13} s^2_{12} \right) 
\cos \frac{ \Delta_{32} x }{2} 
- J_r \sin \delta 
\sin \frac{ \Delta_{32} x }{2} 
\biggr] 
\sin \frac{ \Delta_{21} x }{2} \sin \frac{ \Delta_{31} x }{2}.
\label{decomposed-probability-mu-e}
\end{eqnarray}
In the $\nu_{\mu} \rightarrow \nu_{\tau}$ channel, the decomposed amplitudes and the probabilities can similarly be given by 
\begin{eqnarray} 
S^{ \text{atm} }_{\tau \mu} 
&=&
c_{23} s_{23} c^2_{13} 
\left( e^{ - i \Delta_{31} x} - 1 \right), 
\nonumber \\ 
S^{ \text{sol} }_{\tau \mu} 
&=& 
- \left[ c_{23} s_{23} 
\left( c^2_{12} - s^2_{13} s^2_{12} \right) 
+ s_{13} c_{12} s_{12} 
\left( \cos 2\theta_{23} \cos \delta + i \sin \delta \right)
\right]
\left( e^{ - i \Delta_{21} x} - 1 \right), 
\nonumber \\ 
\label{Smutau-atm-sol-vacuum} 
\end{eqnarray}
and
\begin{eqnarray} 
&& P(\nu_{\mu} \rightarrow \nu_{\tau})^{ \text{non-int-fer} } 
\equiv 
\vert S_{\tau \mu}^{ \text{atm} } \vert^2 
+ \vert S_{\tau \mu}^{ \text{sol} } \vert^2 
= 
c^4_{13} \sin^2 2\theta_{23} 
\sin^2 \frac{ \Delta_{31} x }{2} 
\nonumber \\ 
&+&
\biggl[ 
\left( c^2_{12} - s^2_{13} s^2_{12} \right)^2 
+ s^2_{13} \sin^2 2\theta_{12} 
- \left\{ \cos 2\theta_{23} \left( c^2_{12} - s^2_{13} s^2_{12} \right) 
- 4 J_{rs} \cos \delta \right\}^2 
\biggr]
\sin^2 \frac{ \Delta_{21} x }{2}, 
\nonumber \\
&&
P(\nu_{\mu} \rightarrow \nu_{\tau})^{ \text{int-fer} } 
\equiv 
2 \mbox{Re} \left[ \left( S_{\tau \mu}^{ \text{atm} } \right)^* S_{\tau \mu}^{ \text{sol} } \right] 
\nonumber \\ 
&=& 
8 \biggl[ 
- \left\{
c^2_{13} c^2_{23} s^2_{23} \left( c^2_{12} - s^2_{13} s^2_{12} \right) 
+ \cos 2\theta_{23} J_r \cos \delta 
\right\} \cos \frac{ \Delta_{32} x }{2} 
+ J_r \sin \delta 
\sin \frac{ \Delta_{32} x }{2} 
\biggr]
\sin \frac{ \Delta_{31} x }{2} \sin \frac{ \Delta_{21} x }{2}.
\nonumber \\ 
\label{decomposed-probability-mu-tau}
\end{eqnarray}
In eqs.~\eqref{decomposed-probability-mu-e} and~\eqref{decomposed-probability-mu-tau} we have introduced the simplified notations 
\begin{eqnarray}
&& 
J_r \equiv c_{23} s_{23} c_{12} s_{12} c^2_{13} s_{13}, 
\nonumber \\ 
&& 
J_{rs} \equiv c_{23} s_{23} c_{12} s_{12} s_{13}. 
\label{Jr-def}
\end{eqnarray}
where the former denotes the reduced Jarlskog factor \cite{Jarlskog:1985ht}.
In the $\nu_{\mu} \rightarrow \nu_{\mu}$ disappearance channel, the similar amplitude decomposition leads to
\begin{eqnarray} 
&& P(\nu_{\mu} \rightarrow \nu_{\mu})^{ \text{non-int-fer} }
\equiv 
1 + \vert S_{\mu \mu}^{ \text{atm} } \vert^2 
+ \vert S_{\mu \mu}^{ \text{sol} } \vert^2 
+ 2 \mbox{Re} \left[ S_{\mu \mu}^{ \text{atm} } + S_{\mu \mu}^{ \text{sol} }\right] 
\nonumber \\ 
&=& 
1 - 4 s^2_{23} c^2_{13} \left( 1 - s^2_{23} c^2_{13} \right)
\sin^2 \frac{ \Delta_{31} x }{2} 
\nonumber \\ 
&-& 
4 \left( c^2_{23} c^2_{12} + s^2_{23} s^2_{13} s^2_{12} - 2 J_{rs} \cos \delta \right) 
\left[ c^2_{23} s^2_{12} + s^2_{23} ( 1 - s^2_{13} s^2_{12} ) + 2 J_{rs} \cos \delta \right] 
\sin^2 \frac{ \Delta_{21} x }{2}, 
\nonumber \\ 
&& 
P(\nu_{\mu} \rightarrow \nu_{\mu})^{ \text{int-fer} }
\equiv 
2 \mbox{Re} \left[ \left( S_{\mu \mu}^{ \text{atm} } \right)^* S_{\mu \mu}^{ \text{sol} } \right] 
\nonumber \\ 
&=& 
8 s^2_{23} c^2_{13} 
\left[ c^2_{23} c^2_{12} 
+ s^2_{23} s^2_{13} s^2_{12} 
- 2 J_{rs} \cos \delta \right] 
\cos \frac{ \Delta_{32} x }{2} \sin \frac{ \Delta_{31} x }{2} 
\sin \frac{ \Delta_{21} x }{2}. 
\label{decomposed-probability-mu-mu}
\end{eqnarray}

In the $\nu_{\mu} \rightarrow \nu_{e}$ channel, the dominant component of the interference term is the $\delta$ dependent term, as the $\delta$ independent terms have an extra $s_{13}$ suppression~\cite{Huber:2019frh}. Therefore, one may say that observing the interference term is nearly equivalent of observing the CP phase effect. In the $\nu_{\mu} \rightarrow \nu_{\tau}$ and $\nu_{\mu} \rightarrow \nu_{\mu}$ channels, however, it is not true. There exist the $\delta$ independent pieces in the interference term which have no $s_{13}$ suppression. Therefore, the nature of the interference term, in particular the CP phase dominance or not, depends very much on which channels we discuss, the $\nu_{\mu} \rightarrow \nu_{e}$ channel or the ones in the $\nu_\mu - \nu_\tau$ sector, $\nu_{\mu} \rightarrow \nu_{\tau}$ and $\nu_{\mu} \rightarrow \nu_{\mu}$. We will see in section~\ref{sec:probability-helio-P} that this feature prevails in matter.
 
In the context of discussion above, $\nu_{e}$ and $\bar{\nu}_{e}$ disappearance channels are special with no chance of the probability being $\delta$ dependent even in matter with varying density \cite{Kuo:1987km,Minakata:1999ze}. For this reason the reactor neutrino analysis provides the cleanest place for discussion of nature of the interference term, as stressed in ref.~\cite{Huber:2019frh}. 

\subsection{How to observe the interference term?}
\label{sec:analysis-method}

When the oscillation probability is written as a sum of the interference and the non-interference terms, $P(\nu_{\beta} \rightarrow \nu_{\alpha}) = P(\nu_{\beta} \rightarrow \nu_{\alpha})^{ \text{non-int-fer} } + P(\nu_{\beta} \rightarrow \nu_{\alpha})^{ \text{int-fer} }$, one can design a simple $\chi^2$ test to know at what significance level one observes existence of the interference effect \cite{Huber:2019frh}. 
To quantify the statistical significance, we define the test probability by introducing the $q$ parameter 
\begin{eqnarray} 
P(\nu_{\beta} \rightarrow \nu_{\alpha}) 
&=& 
P(\nu_{\beta} \rightarrow \nu_{\alpha})^{ \text{non-int-fer} } 
+ q P(\nu_{\beta} \rightarrow \nu_{\alpha})^{ \text{int-fer} }. 
\label{ansatz}
\end{eqnarray}
We calculate $\chi^2 (q)$ by fitting the data with the ansatz \eqref{ansatz} with marginalization over the standard oscillation parameters including $\delta$. 
The $\chi^2 (q)$ has one degree of freedom, and is expected to have a minimum at $q=1$. Depending upon how deep is the minimum, we can make statement on at what CL one observes the quantum interference between the atmospheric and the solar amplitudes. 
This procedure is employed in the analysis of JUNO-like setting but, of course, without marginalization over $\delta$ \cite{Huber:2019frh}. Since the structure of the probability written by the decomposed components with the $q$ extension is universal, we expect that the analysis procedure with eq.~\eqref{ansatz} applies to all the flavor channels. 

\section{Amplitude decomposition in matter: Problems and our approach}
\label{sec:approach}

Since extension of the amplitude decomposition to an environment in matter will reveal a highly nontrivial feature we first explain, in words, what are the problems and our approach to resolve them. The readers may find in this section a rough sketch of the design plan for this paper. 

Let $\lambda_{i}$ ($i=1,2,3$) be the eigenvalues of $2E H$ and $V$ the unitary matrix which diagonalizes the Hamiltonian $H$. In matter, $\lambda_{i}$ and the mixing matrix $V$, both of which depend on the matter potential, replace $m^2_{i}$ and $U_{\text{\tiny MNS}}$ matrix, respectively, in vacuum. Then, one can define the amplitude decomposition in matter by elevating the eigenvalues and the mixing matrix into those in matter, $m^2_{i} \rightarrow \lambda_{i}$ and $U \rightarrow V$, in eq.~\eqref{atm-sol-amplitude-vac}. See section~\ref{sec:principle} for more details. The procedure will allow us to define the amplitude decomposition in matter which is exactly parallel to eq.~\eqref{atm-sol-amplitude-vac} in vacuum. The exact expressions of $\lambda_{i}$ and $V$ are known under the uniform matter density approximation~\cite{Zaglauer:1988gz}, and hence this method may be called as the Zaglauer-Schwarzer (ZS) decomposition. 

However, what is nontrivial is the interpretation of the ZS decomposition. In vacuum, the atmospheric and the solar waves are defined as the $\Delta m^2_{31}$-driven and $\Delta m^2_{21}$-driven oscillations, respectively \cite{Huber:2019frh}. In taking the ZS decomposition, it is natural to assume that the ``atmospheric'' and the ``solar'' waves in matter are defined by the frequencies determined by $\lambda_{3} - \lambda_{2}$ in the normal ($\lambda_{3} - \lambda_{1}$ in the inverted) mass ordering and $\lambda_{2} - \lambda_{1}$, respectively. It may work in region where modification of the eigenvalues by the matter effect is modest. But, it is known that the eigenvalues $\lambda_{i}$ become dynamical at high energies or high matter densities, and the difference $\lambda_{2} - \lambda_{1}$ for the ``solar'' oscillation can be much larger than the  ``atmospheric'' energy splitting in certain region of kinematical phase space.\footnote{
%%%%%%%%%%%%% footnote %%%%%%%%%%%%%%%%
It is high-energy or high-density region $| Y_{e} \rho E | \gg 20 \mbox{g~cm}^{-3}$ 
in the anti-neutrino (neutrino) channel in the normal (inverted) mass ordering, where $\rho$ denotes the matter density, $Y_{e}$ is the number of electron per nucleon. 
}
Is it still possible to interpret $\lambda_{2} - \lambda_{1}$ wave as the ``solar'' oscillation in this region? Which property does really define the wave is either the ``solar'' wave, or the ``atmospheric'' wave?

Since we do not know the general, precise answer to these questions we take another approach in this paper. We restrict ourselves into the region where we know how the atmospheric and the solar waves are modified by the matter effect. In regions of the atmospheric-scale and the solar-scale enhanced oscillations the appropriate perturbative frameworks are formulated which can serve for this purpose. In a sense, we take a ``bottom-up'' approach by analyzing these theories to learn what is the right way of decomposing the oscillation $S$ matrix into the ``atmospheric'' and the ``solar'' amplitudes in matter.

In fact, the matter-effect modification of the eigenvalues is not the whole issue. Even with infinitesimal matter potential there is a mode of oscillation whose nature can only be described as inherent mixture of the atmospheric and the solar waves. In this case the eigenvalues are approximately the same as in vacuum. Yet, the presence of such mixed wave prevents us from using the general definition 1 and 2 of the atmospheric and solar amplitudes given in section~\ref{sec:amplitude-decomposition-vacuum}. See section~\ref{sec:decomposition-small-matter} for discussion of this point. Therefore, it appears to us that the conceptual issues are immanent in the un-understood aspects of the amplitude decomposition in matter.

Finally, we note, in spite of the above comments, that the ZS decomposition will play an important role in the amplitude decomposition in matter. 

\section{The three-flavor neutrino evolution in matter} 
\label{sec:3nu-evolution }

First, we define the system of three-flavor neutrino evolution in matter. Though standard and well known, we do it to define notations. 
The evolution of the three-flavor neutrinos in matter can be described by the Schr\"odinger equation in the flavor basis, $i \frac{d}{dx} \nu = H \nu$, with Hamiltonian 
\begin{eqnarray}
H= 
\frac{ 1 }{ 2E } \left\{ 
U \left[
\begin{array}{ccc}
0 & 0 & 0 \\
0 & \Delta m^2_{21}& 0 \\
0 & 0 & \Delta m^2_{31} 
\end{array}
\right] U^{\dagger}
+
\left[
\begin{array}{ccc}
a(x) & 0 & 0 \\
0 & 0 & 0 \\
0 & 0 & 0
\end{array}
\right] 
\right\}, 
\label{flavor-basis-hamiltonian}
\end{eqnarray}
where $E$ is neutrino energy and $\Delta m^2_{ji} \equiv m^2_{j} - m^2_{i}$. 
In (\ref{flavor-basis-hamiltonian}), $U \equiv U_{\text{\tiny MNS}}$ denotes the standard $3 \times 3$ lepton flavor mixing matrix~\cite{Maki:1962mu} which relates the flavor neutrino states to the vacuum mass eigenstates as $\nu_{\alpha} = U_{\alpha i} \nu_{i}$, where $\alpha$ runs over $e, \mu, \tau$, and the mass eigenstate index $i$ runs over $1,2,$ and $3$. 
We use, except for in section~\ref{sec:DMP}, the lepton flavor mixing matrix in the PDG convention \cite{Tanabashi:2018oca} 
\begin{eqnarray}
U_{\text{\tiny PDG}} 
&=& 
\left[
\begin{array}{ccc}
1 & 0 &  0  \\
0 & c_{23} & s_{23} \\
0 & - s_{23} & c_{23} \\
\end{array}
\right] 
\left[
\begin{array}{ccc}
c_{13}  & 0 &  s_{13} e^{- i \delta} \\
0 & 1 & 0 \\
- s_{13} e^{ i \delta} & 0 & c_{13}  \\
\end{array}
\right] 
\left[
\begin{array}{ccc}
c_{12} & s_{12}  &  0  \\
- s_{12} & c_{12} & 0 \\
0 & 0 & 1 \\
\end{array}
\right] 
\equiv 
U_{23} U_{13} U_{12} 
\nonumber \\
&=& 
\left[
\begin{array}{ccc}
c_{12} c_{13} & s_{12} c_{13} & s_{13} e^{-i\delta} \\
- s_{12} c_{23} - c_{12} s_{23} s_{13} e^{i\delta} &
c_{12} c_{23} - s_{12} s_{23} s_{13} e^{i\delta} & s_{23} c_{13} \\
s_{12} s_{23} - c_{12} c_{23} s_{13} e^{i\delta} &
- c_{12} s_{23} - s_{12} c_{23} s_{13} e^{i\delta} & c_{23} c_{13} \\
\end{array}
\right].
\label{MNSmatrix-PDG}
\end{eqnarray}
The functions $a(x)$ in (\ref{flavor-basis-hamiltonian}) denote the Wolfenstein matter potential \cite{Wolfenstein:1977ue} due to charged current (CC) reactions 
\begin{eqnarray} 
a &=&  
2 \sqrt{2} G_F N_e E \approx 1.52 \times 10^{-4} \left( \frac{Y_e \rho}{\rm g\,cm^{-3}} \right) \left( \frac{E}{\rm GeV} \right) {\rm eV}^2.  
\label{matt-potential}
\end{eqnarray}
Here, $G_F$ is the Fermi constant, $N_e$ is the electron number density in matter. $\rho$ and $Y_e$ denote, respectively, the matter density and number of electron per nucleon in matter. For simplicity and clarity we will work with the uniform matter density approximation throughout this paper. But, it is not difficult to extend our treatment to varying matter density case if adiabaticity holds.

\section{Amplitude decomposition with infinitesimal matter potential} 
\label{sec:decomposition-small-matter}

Since we know how to decompose the $S$ matrix into the atmospheric and solar amplitudes in vacuum, a natural first step is to introduce the matter potential with a tiny magnitude. Then, the system can be analyzed perturbatively. The framework, so called the matter perturbation theory,\footnote{
%%%%%%%%%%%%% footnote %%%%%%%%%%%%%
In this paper, we mean by the ``matter perturbation theory'' a perturbative framework with the unique expansion parameter $a / \Delta m^2_{31}$ without introducing any further approximations. } 
is known since the early era, see e.g., \cite{Barger:1980tf,Minakata:1998bf}. As its formulation is well known we just sketch out the formalism in appendix~\ref{sec:matter-P-theory}. We briefly mention here that we use the vacuum mass-eigenstate basis with Hamiltonian $\check{H} = \left( U_{23} U_{13} U_{12} \right)^{\dagger} H U_{23} U_{13} U_{12}$ to define the perturbation theory, and treat the first and the second terms in eq.~\eqref{flavor-basis-hamiltonian} as the unperturbed and perturbed parts of the Hamiltonian, respectively. The zeroth-order eigenvalues of the Hamiltonian are denoted as $h_{i}$ ($i=1,2,3$) and they are given by 
\begin{eqnarray} 
h_{1} = 0, 
\hspace{10mm}
h_{2} = \Delta_{21}, 
\hspace{10mm}
h_{3} = \Delta_{31}. 
\label{eigenvalues-matter-P}
\end{eqnarray}
For convenience, we introduce another simplified notation 
\begin{eqnarray} 
\Delta_{a} \equiv \frac{ a }{ 2E },
\label{Delta-a-def}
\end{eqnarray}
in addition to $\Delta_{ji} \equiv \Delta m^2_{ji} / 2E$ in eq.~\eqref{Delta-ij-def}.

\subsection{$S$ matrix in the flavor basis: $\nu_\mu \rightarrow \nu_e$ channel }
\label{sec:matter-P-mu-e}

In appendix~\ref{sec:matter-P-theory}, we compute all the $\check{S}$ matrix elements in the vacuum mass eigenstate basis to first order in the matter perturbation theory. Then, the flavor basis $S$ matrix can be obtained as $S = U \check{S} U^{\dagger}$.
In this paper we focus on the $\nu_\mu \rightarrow \nu_e$ channel in most cases to examine how the decomposition of the $S$ matrix elements into the atmospheric and solar amplitudes can be (or cannot be) done in matter. It is because our concern is primarily on the conceptual issue on how the decomposition can be performed correctly. Since the zeroth-order $S$ matrix is identical with the one in vacuum, we discuss the first-order term. 

Using the eigenvalues given in \eqref{eigenvalues-matter-P}, the relevant matrix elements can be calculated in first order in the matter perturbation theory, see eqs.~\eqref{check-Smatrix} and \eqref{flavor-Smatrix}. By using them, we obtain the flavor basis $S$ matrix element $S_{e \mu}^{(1)}$ as 
\begin{eqnarray}
S_{e \mu}^{(1)}
&=& 
c_{12} s_{12} c^3_{13} 
\left( \cos 2\theta_{12} c_{23} - \sin 2\theta_{12} s_{13} s_{23} e^{ - i \delta} \right) 
\frac{ \Delta_{a} }{ \Delta_{21} } 
\left\{ e^{ - i \Delta_{21} x } - 1 \right\} 
\nonumber \\
&+& 
c_{12} c_{13} s_{13} 
\left( - s_{12} s_{13} c_{23} + \cos 2\theta_{13} c_{12} s_{23} e^{ -i \delta }  \right) 
\frac{ \Delta_{a} }{ \Delta_{31} } 
\left\{ e^{ - i \Delta_{31} x } - 1 \right\} 
\nonumber \\
&+& 
s_{12} c_{13} s_{13} 
\left( c_{12} s_{13} c_{23} + \cos 2\theta_{13} s_{12} s_{23} e^{ -i \delta } \right) 
\frac{ \Delta_{a} }{ \Delta_{31} - \Delta_{21} } 
\left\{ e^{ - i \Delta_{31} x } - e^{ - i \Delta_{21} x } \right\} 
\nonumber \\
&+& 
( -i \Delta_{a} x) 
\biggl[ 
c^3_{12} c^3_{13} 
\left( - s_{12} c_{23} - c_{12} s_{13} s_{23} e^{ - i \delta} \right) 
\nonumber \\
&+& 
s^3_{12} c^3_{13} 
\left( c_{12} c_{23} - s_{12} s_{13} s_{23} e^{- i \delta}  \right)
e^{ - i \Delta_{21} x } 
+ c_{13} s^3_{13} s_{23} e^{-i\delta} 
e^{ - i \Delta_{31} x } 
\biggr]. 
\label{Semu-matter-P-1st}
\end{eqnarray}
Now, with $S_{e \mu}^{(1)}$ in \eqref{Semu-matter-P-1st} at hand, one can apply the definition 1 and 2 of the atmospheric and solar amplitudes, eqs.~\eqref{atm-sol-amplitude-vac-def} and \eqref{completeness}, given in the previous section~\ref{sec:decomposition-vacuum}. One immediately notices that it fails. One obtains $S_{\alpha \beta}^{ \text{atm} }$ and $S_{\alpha \beta}^{ \text{sol} }$ by taking the limits $\Delta_{21} \rightarrow 0$ and $\Delta_{31} \rightarrow 0$, respectively, as 
\begin{eqnarray}
&& 
\left( S_{e \mu}^{ \text{atm} } \right)^{(1)} = 
c_{13} s_{13} s_{23} e^{ - i \delta} 
\left[
\left\{ - \cos 2\theta_{13} 
+ s^2_{13} \left( e^{ - i \Delta_{31} x } - 1 \right) 
\right\} ( -i \Delta_{a} x) 
+ 
\cos 2\theta_{13} 
\frac{ \Delta_{a} }{ \Delta_{31} } 
\left( e^{ - i \Delta_{31} x } - 1 \right)
\right], 
\nonumber \\
&&
\left( S_{e \mu}^{ \text{sol} } \right)^{(1)} =
s_{12} c_{13} 
\left( c_{12} c_{23} - s_{12} s_{13} s_{23} e^{- i \delta} \right)
\nonumber \\
&\times& 
\biggl[
- \left\{ ( 1 - 2 s^2_{12} c^2_{13} ) 
- s^2_{12} c^2_{13} \left( e^{ - i \Delta_{21} x } - 1 \right) 
\right\} ( -i \Delta_{a} x) 
+ 
\left( 1 - 2 s^2_{12} c^2_{13} \right) 
\frac{ \Delta_{a} }{ \Delta_{21} } 
\left( e^{ - i \Delta_{21} x } - 1 \right) ]
\biggr],
\end{eqnarray}
which satisfy the consistency conditions $\lim_{\Delta_{31} \rightarrow 0} S_{\alpha \beta}^{ \text{atm} } 
= \lim_{\Delta_{21} \rightarrow 0} S_{\alpha \beta}^{ \text{sol} } = 0$. But, the second condition, the completeness condition, cannot be met.

The cause of the problem is obvious, the third term in \eqref{Semu-matter-P-1st}. Before inserting the PDG expression of the $U$ matrix elements it took the form 
\begin{eqnarray}
&& 
\left( U_{e3} U_{\mu 2}^* e^{ i \delta } + U_{e2} U_{\mu 3}^* e^{ -i \delta } \right)
s_{12} c_{13} s_{13} 
\frac{ \Delta_{a} }{ h_{3} - h_{2} } 
\left( e^{ - i h_{3} x } - e^{ - i h_{2} x } \right). 
\label{3rd-term}
\end{eqnarray}
A diagrammatical representation of this term would consist of the two amplitudes which describe perturbative transition via $H^{(1)}_{23}$ or $H^{(1)}_{32}$:
\begin{eqnarray}
&& 
\nu_{\mu} - U_{\mu 2}^* - \nu_{2} \rightarrow H^{(1)}_{23} - \nu_{3} \rightarrow U_{e 3} - \nu_{e}, 
\nonumber \\
&& 
\nu_{\mu} - U_{\mu 3}^* - \nu_{3} \rightarrow H^{(1)}_{32} - \nu_{2} \rightarrow U_{e 2} - \nu_{e}.
\label{diagram-structure}
\end{eqnarray}
They are the genuine mixed effect of both the $\Delta_{31}$- and $\Delta_{21}$-driven waves, and therefore they cannot be decomposed to the pure atmospheric and the pure solar amplitudes in the manner that was possible in vacuum.

Notice that $\frac{1}{(\Delta_{31} - \Delta_{21})} = \frac{1}{\Delta_{32}}$ so that the structure of the first three lines in eq.~\eqref{Semu-matter-P-1st} is natural with the three possible forms of the energy denominators. Therefore, it appears that the problem is caused by the rigid definition of amplitude decomposition, not by $S_{e \mu}^{(1)}$ itself. Despite we see no green light for our narrow definition of amplitude decomposition, eqs.~\eqref{atm-sol-amplitude-vac-def} and \eqref{completeness}, to survive in matter, we ask the question for a complete understanding: Is there any case in which our vacuum definition is valid in matter? In the next section we find the answer is ``Yes''.

\section{Amplitude decomposition in the helio-matter perturbation theory} 
\label{sec:AKS}

In looking for the principle of decomposing the neutrino oscillation $S$ matrix to the atmospheric and the solar amplitudes in neutrino oscillation in matter, we examine one of the simplest perturbative frameworks discussed by Arafune, Koike, and Sato (AKS) \cite{Arafune:1997hd}, which may be called as the ``helio-matter perturbation theory''. We assume that we are around the atmospheric-scale enhanced oscillation and regard $\Delta_{21} / \Delta_{31} = \Delta m^2_{21} / \Delta m^2_{31}$ as well as $\Delta_{a} / \Delta_{31} = a / \Delta m^2_{31}$ the small expansion parameters. In appendix~\ref{sec:formulation-AKS}, we give a brief description of its formulation using the same vacuum mass-eigenstate basis as in appendix~\ref{sec:matter-P-theory}. 
For clarity of terminology, not to confuse it with the helio perturbation theory to be discussed in section~\ref{sec:helio-perturbation-theory}, we will call the framework the AKS perturbation theory in this paper. 

\subsection{$S$ matrix and the amplitude decomposition: $\nu_\mu \rightarrow \nu_e$ channel}
\label{sec:amplitude-decomposition-AKS}

The flavor basis $S$ matrix elements can be calculated by using eq.~\eqref{S-matter-1st} with the vacuum mass-eigenstate basis elements in eq.~\eqref{check-Smatrix-1st}. For the purpose of our discussion, we compute here the first order corrections. There are two terms, $\left( S^{(1)}_{ \text{matter} } \right)_{e \mu}$ and $\left( S^{(1)}_{ \text{helio} } \right)_{e \mu}$. A straightforward calculation leads to 
\begin{eqnarray}
&& \left( S^{(1)}_{ \text{matter} } \right)_{e \mu} = 
%\nonumber \\&=& 
U_{12} U_{21}^* c_{12} s_{12} c^2_{13} (-i \Delta_{a} x) + U_{11} U_{22}^* c_{12} s_{12} c^2_{13} (-i \Delta_{a} x) 
\nonumber \\
&+& 
U_{11} U_{21}^* c^2_{12} c^2_{13} (-i \Delta_{a} x) + U_{12} U_{22}^* s^2_{12} c^2_{13} (-i \Delta_{a} x) + U_{13} U_{23}^* s^2_{13} (-i \Delta_{a} x) e^{ -i \Delta_{31} x } 
\nonumber \\
&-&
U_{13} U_{21}^* c_{12} c_{13} s_{13} e^{ i \delta } \Delta_{a} 
\frac{ 1 - e^{ -i \Delta_{31} x } }{ \Delta_{31} } 
- U_{11} U_{23}^* c_{12}c_{13} s_{13} e^{ -i \delta } \Delta_{a} 
\frac{ 1 - e^{ -i \Delta_{31} x } }{ \Delta_{31} } 
\nonumber \\
&-&
U_{13} U_{22}^* s_{12} c_{13} s_{13} e^{ i \delta } \Delta_{a} 
\frac{ 1 - e^{ -i \Delta_{31} x } }{ \Delta_{31} } 
- U_{12} U_{23}^* s_{12} c_{13} s_{13} e^{ -i \delta } \Delta_{a} 
\frac{ 1 - e^{ -i \Delta_{31} x } }{ \Delta_{31} }. 
\label{S-1st-matter-AKS-1}
\end{eqnarray}
But, a simplification occurs and $\left( S^{(1)}_{ \text{matter} } \right)_{e \mu}$ has a simpler expression\footnote{
%%%%%%%%%%%% footnote %%%%%%%%%%%%
We note that while showing disappearance of $\theta_{12}$ dependence explicitly in this way is pedagogical, the simplest way of recognizing this feature is to use the basis $\hat{H} = U_{13}^{\dagger} U_{23}^{\dagger} H U_{23} U_{13}$, which will  be used in appendix~\ref{sec:formulation-helio-perturbation}. See eq.~\eqref{tilde-H-def}. It is the most convenient basis for the AKS perturbation theory. 
} 
\begin{eqnarray}
&& \left( S^{(1)}_{ \text{matter} } \right)_{e \mu} = 
%\nonumber \\&=& 
- c_{13} s_{13} s_{23} e^{ - i \delta} (-i \Delta_{a} x) 
\left[ c^2_{13} - s^2_{13} e^{ -i \Delta_{31} x } \right] 
- \cos 2\theta_{13}
c_{13} s_{13} s_{23} e^{ -i \delta } 
\frac{  \Delta_{a} }{ \Delta_{31} } 
\left( 1 - e^{ -i \Delta_{31} x } \right). 
\nonumber \\
\label{S-1st-matter-AKS-2}
\end{eqnarray}
Notice that all the $\theta_{12}$ dependence disappeared in going from \eqref{S-1st-matter-AKS-1} to \eqref{S-1st-matter-AKS-2}. Furthermore, $\left( S^{(1)}_{ \text{matter} } \right)_{e \mu}$ consists only of $\Delta_{31}$ with desirable property that it vanishes as $\Delta_{31} \rightarrow 0$. Therefore, almost certainly $\left( S^{(1)}_{ \text{matter} } \right)_{e \mu}$ contributes purely to the atmospheric amplitude. The zeroth order element $S^{(0)}_{e \mu} =  U_{13} U_{23}^* \left( e^{ -i \Delta_{31} x } - 1 \right) = c_{13} s_{13} s_{23} e^{-i \delta} \left( e^{ -i \Delta_{31} x } - 1 \right)$ is also the atmospheric amplitude. 
On the other hand, the helio correction
\begin{eqnarray} 
&& 
\left( S^{(1)}_{ \text{helio} } \right)_{e \mu} 
%\nonumber \\&=& 
= ( - i \Delta_{21} x) U_{12} U_{22}^* 
= ( - i \Delta_{21} x) 
s_{12} c_{13} \left( c_{12} c_{23} - s_{12} s_{23} s_{13} e^{ - i \delta} \right),  
\label{S-1st-helio-AKS}
\end{eqnarray}
which depends only on $\Delta_{21}$ and vanishes in the $\Delta_{21} \rightarrow 0$ limit, must be the solar amplitude. 

Therefore, the oscillation amplitude $S_{e \mu}$ to first order in the AKS expansion can be decomposed into the atmospheric and the solar amplitudes as $S_{e \mu} = S_{e \mu}^{ \text{atm} } + S_{e \mu}^{ \text{sol} }$ where 
\begin{eqnarray} 
S_{e \mu}^{ \text{atm} } &=& 
- e^{ - i \delta} s_{23} c_{13} s_{13} 
\biggl[
\left( 1 - e^{ -i \Delta_{31} x } \right) 
+ (-i \Delta_{a} x) 
\left[ c^2_{13} - s^2_{13} e^{ -i \Delta_{31} x } \right] 
+ \cos 2\theta_{13}
\frac{  \Delta_{a} }{ \Delta_{31} } 
\left( 1 - e^{ -i \Delta_{31} x } \right) 
\biggr], 
\nonumber \\
S_{e \mu}^{ \text{sol} } &=&
( - i \Delta_{21} x) 
s_{12} c_{13} \left( c_{12} c_{23} - s_{12} s_{23} s_{13} e^{ - i \delta} \right). 
\label{atm-sol-amplitude-vac-AKS}
\end{eqnarray}
It can be easily checked that they satisfy the definition of amplitude decomposition of narrow sense given in section~\ref{sec:decomposition-vacuum}, $\lim_{\Delta_{21} \rightarrow 0} \left( S_{e \mu} ^{ \text{atm} } + S_{e \mu}^{ \text{sol} } \right) = S_{e \mu} ^{ \text{atm} }$ and $\lim_{\Delta_{31} \rightarrow 0} \left( S_{e \mu} ^{ \text{atm} } + S_{e \mu}^{ \text{sol} } \right) = S_{e \mu} ^{ \text{sol} }$, with the consistency condition $\lim_{ \Delta_{31} \rightarrow 0} S_{e \mu}^{ \text{atm} } =0$, and the one for $S_{e \mu}^{ \text{solar} }$. Notice that the completeness condition must be satisfied to first order because no terms has been dropped during the process to reach \eqref{atm-sol-amplitude-vac-AKS}. They have a vacuum limit which agrees with the one in section~\ref{sec:decomposition-vacuum}. Thus, the vacuum definition of amplitude decomposition works to first order in the AKS helio-matter perturbation theory.

In fact, the expression of $S_{e \mu}^{ \text{sol} }$ in \eqref{atm-sol-amplitude-vac-AKS} is akin to the one in vacuum, see eq.~\eqref{Semu-atm-sol-vacuum}. Since $\Delta_{21} x \ll 1$ in region of applicability of the AKS framework the factor $( - i \Delta_{21} x)$ can be understood as $\left( e^{ - i \Delta_{21} x } -1 \right)$ in an excellent approximation. Therefore, the matter-effect modification in the decomposed amplitudes exists essentially only in the atmospheric amplitude $S_{e \mu}^{ \text{atm} }$.

\subsection{The oscillation probability $P(\nu_\mu \rightarrow \nu_e)$: AKS}

The oscillation probability $P(\nu_\mu \rightarrow \nu_e)$ is given to first order in the AKS expansion as 
\begin{eqnarray} 
P(\nu_\mu \rightarrow \nu_e) 
= 
\vert S_{e \mu} ^{ \text{atm} } + S_{e \mu}^{ \text{sol} } \vert^2 
= 
\vert S_{e \mu} ^{ \text{atm} } \vert^2 
+ 2 \mbox{Re} 
\left[ \left( S_{e \mu}^{ \text{atm} } \right)^* S_{e \mu}^{ \text{sol} } \right]. 
\label{P-mue-AKS}
\end{eqnarray}
The atmospheric term (the first term in \eqref{P-mue-AKS}) is given, ignoring the second order $(\Delta_{a}/\Delta_{31})^2$ terms, by 
\begin{eqnarray} 
&& 
P(\nu_{\mu} \rightarrow \nu_{e})^{ \text {non-int-fer} } 
= \vert S_{e \mu} ^{ \text{atm} } \vert^2
\nonumber \\ 
&=& 
s^2_{23} c^2_{13} s^2_{13} 
\biggl[ 
4 \left( 
1 + 2 \cos 2\theta_{13} \frac{ \Delta_{a} }{ \Delta_{31} } 
\right) 
\sin^2 \frac{ \Delta_{31} x }{2} 
- 2 \cos 2\theta_{13} ( \Delta_{a} x ) \sin \Delta_{31} x 
\biggr].
\label{P-mue-ampsq:AKS}
\end{eqnarray}
The interference term, the term of our interest, is given by 
\begin{eqnarray} 
&& P(\nu_\mu \rightarrow \nu_e)^{ \text {int-fer} } 
= 2 \mbox{Re} 
\left[ \left( S_{e \mu}^{ \text{atm} } \right)^* S_{e \mu}^{ \text{sol} } \right] 
\nonumber \\
&=& 
2 J_{r} ( \Delta_{21} x ) 
\left[ \cos \delta \sin \Delta_{31} x - 2 \sin \delta \sin^2 \frac{ \Delta_{31} x }{2}
\right] 
- 2 s^2_{23} c^2_{13} s^2_{13} s^2_{12} 
( \Delta_{21} x ) \sin \Delta_{31} x,
%\nonumber \\
\label{P-mue-interfere:AKS}
\end{eqnarray}
where the reduced Jarlskog factor $J_r = c_{23} s_{23} c_{12} s_{12} c^2_{13} s_{13}$ is defined in \eqref{Jr-def}. 

\subsection{AKS decomposition: Unique successful case?}  

Thus, we have found a concrete example in which the vacuum definition of the amplitude decomposition, eqs.~\eqref{atm-sol-amplitude-vac-def} and \eqref{completeness}, works in matter. It may be applicable at low energy of $\sim$ several 100 MeV and medium baseline of a few $\times 100$ km, under the given hierarchy of the two $\Delta m^2$, $\epsilon \equiv \Delta m^2_{21} / \Delta m^2_{31} \ll 1$. We note that the region nicely matches the setting of the T2K \cite{Abe:2019vii}, T2HK \cite{Abe:2018uyc}, and ESS$\nu$SB \cite{Baussan:2013zcy} experiments, which may be called as the ``cleanest region'' for the amplitude decomposition in matter because the decomposed waves retain the original frequencies associated with the $\Delta m^2_{31}$ and $\Delta m^2_{21}$ as in vacuum.\footnote{
%%%%%%%%%%%%% footnote %%%%%%%%%%%%%%
It is nice to see that the region, which was proposed for clean measurement of CP phase $\delta$ with minimal matter effect by the low-energy mu-neutrino superbeam \cite{Minakata:2000ee}, also reveals the favorable feature for the amplitude decomposition with the vacuum frequencies. 
}

Now, one may ask: Why is the AKS framework successful while the matter perturbation not, despite that the both expand to first order? The answer is that the troublesome aspect of the ``third term'' in eq.~\eqref{Semu-matter-P-1st} goes away because we can approximate the energy denominator as $\frac{1}{(\Delta_{31} - \Delta_{21})} \simeq \frac{1}{\Delta_{31}}$, because it is inside the matter-suppressed first order term. In consistent with this observation, we have checked that the validity of the vacuum definition does not survive when the second-order AKS corrections are added. 

This understanding suggests that the validity of the vacuum prescription of the amplitude decomposition in matter necessitates the both expansion parameters $\Delta m^2_{21} / \Delta m^2_{31}$ and $a / \Delta m^2_{31}$ be small. The smallness of the matter effect is required because otherwise the eigenvalues become dynamical. Then, it is likely that the first order AKS perturbation theory is the unique case which retains the vacuum definition of the amplitude decomposition. The treatment of the helio perturbation theory with all order effect of matter given in section~\ref{sec:helio-perturbation-theory} will confirm our expectation. Therefore, in matter environment in general, we need to depart from the vacuum definition of the amplitude decomposition. 

\section{Principle of decomposition of the $S$ matrix in matter} 
\label{sec:principle}

To make progress, let us summarize the lessons we have learned so far, in particular, from the failure of our definition of amplitude decomposition, eqs.~\eqref{atm-sol-amplitude-vac-def} and \eqref{completeness} in matter. In general, the following two issues are involved.

\begin{itemize}

\item 
The eigenvalues of the Hamiltonian $h_{i}$ (i=1,2,3) of the three mass eigenstates are in general different from the vacuum values $m^2_{i} / 2E$. 

\item
The genuine three-generation structure of the neutrino oscillation produces mixture of the $\Delta m^2_{31}$- and $\Delta m^2_{21}$-driven waves.

\end{itemize}
\noindent 
Under a matter potential whose magnitude are comparable with the vacuum effect, $a \sim \Delta m^2_{31}$, the eigenvalues of the three mass eigenstates can be significantly different from the vacuum values. In this case, the physical meaning of the limiting procedure defined with the vacuum eigenvalues, $\Delta m^2_{31}$ and $\Delta m^2_{21}$, becomes obscure. The justification of the prescription that the $\Delta m^2_{21} \rightarrow 0$ ($\Delta m^2_{31} \rightarrow 0$) limit defines the atmospheric (solar) amplitude may lose the original meaning. 

However, we must note that modification of the eigenvalues is not the whole issue. In the matter perturbation theory, the eigenvalues are the same as in vacuum, see eq.~\eqref{eigenvalues-matter-P}. The failure of our definition is due to the presence of mixed atmospheric- and solar-scale oscillation mode, which is inherent to the three-generation structure of the lepton world.

\subsection{What are the two independent dynamical modes of oscillations}
\label{sec:degree-oscillations} 

Now, we address the principle of amplitude decomposition in matter. Our failure in imposing the vacuum definition of amplitude decomposition (see sections~\ref{sec:decomposition-small-matter} and the following ones) teaches us that the atmospheric and the solar oscillations in the narrow sense are not always the appropriate two independent dynamical degrees of freedom in describing the three-flavor neutrino transformation in matter. The important fact is that it is true even if the matter potential $a$ is much smaller than the vacuum effect $\sim \Delta m^2_{31}$, which testifies that nature of the difficulty is a conceptual one, not technical one, as emphasized above. What we should do is, therefore, 
\begin{itemize}
\item
To identify the appropriate dynamical degrees of freedom, which we call ``A'' and ``S'' in more generic environments. In the vicinity of region of validity of our perturbative formulas, A and S may be the matter-dressed atmospheric and the matter-dressed solar oscillations, respectively. 

\item 
To formulate a systematic way of computing $S^{A}$ and $S^{S}$ for amplitude decomposition, for which the completeness condition $S_{\alpha \beta} = \delta_{\alpha \beta} + S_{\alpha \beta}^{A} + S_{\alpha \beta}^{S}$ is automatically satisfied. 

\end{itemize}
\noindent
It is a highly nontrivial task, and the method for carrying it out systematically for a generic matter density is not known to the present author. 

\subsection{Zaglauer-Schwarzer decomposition}

In fact, the recognition that the atmospheric and the solar waves do not necessarily provide the appropriate two independent dynamical degrees of freedom in the three-flavor oscillation in matter is not new. In an effort to find the exact solution of the three-flavor neutrino evolution in uniform-density matter, Zaglauer and Schwarzer identified them albeit in an abstract fashion \cite{Zaglauer:1988gz}. It is shown that under the uniform matter density approximation the oscillation $S$ matrix can be written exactly in the same form as in vacuum 
\begin{eqnarray} 
S_{\alpha \beta} = 
V_{\alpha 1} V^{*}_{\beta 1} e^{ - i \frac{ \lambda_{1} }{2E} x} 
+ V_{\alpha 2} V^{*}_{\beta 2} e^{ - i \frac{ \lambda_{2} }{2E} x}
+ V_{\alpha 3} V^{*}_{\beta 3} e^{ - i \frac{ \lambda_{3} }{2E} x}. 
\label{S-matrix-matter}
\end{eqnarray}
It can be obtained by the replacements $m^2_{i} \rightarrow \lambda_{i} ~~(i=1,2,3)$ and $U_{\alpha i} \rightarrow V_{\alpha i}$ in eq.~\eqref{S-matrix-def-original}. Here, $\lambda_{i}$ denotes the eigenvalues of $2E H$, where $H$ denotes the Hamiltonian in the flavor basis \eqref{flavor-basis-hamiltonian} but with slightly different phase convention,\footnote{
%%%%%%%%%%%% footnote %%%%%%%%%%%%%%
We consider the slightly different Hamiltonian from \eqref{flavor-basis-hamiltonian} whose vacuum part takes the form $U \text{diag} ( m^2_{1}/2E, m^2_{2}/2E, m^2_{3}/2E ) U^{\dagger}$. 
}
and $V$ is the unitary matrix which diagonalizes the Hamiltonian. The explicit expressions of $\lambda_{i}$ and $V$ are obtained in refs.~\cite{Barger:1980tf} and~\cite{Zaglauer:1988gz}, respectively.

With eq.~\eqref{S-matrix-matter}, the same treatment of amplitude decomposition in vacuum as described in section~\ref{sec:decomposition-vacuum} goes through in matter. Using two different ways of taking the trace of the Hamiltonian one can derive the sum rule (see e.g., ref.~\cite{Kimura:2002wd})
\begin{eqnarray} 
\lambda_{1} + \lambda_{2} + \lambda_{3} 
= m^2_{1} + m^2_{2} + m^2_{3} + a  
\label{sum-rule}
\end{eqnarray}
which tells us that only two out of the three eigenvalues $\lambda_{i}$ are independent. It means that only two amplitudes are independent. One can similarly define the amplitudes $S_{\alpha \beta}^{A}$ and $S_{\alpha \beta}^{S}$ as 
\begin{eqnarray} 
&& 
S_{\alpha \beta}^{A} 
\equiv  
V_{\alpha 3} V^{*}_{\beta 3} 
\left[ e^{ - i \frac{ ( \lambda_{3} - \lambda_{1} ) }{2E} x} - 1 \right], 
\nonumber \\
&& 
S_{\alpha \beta}^{S} \equiv  
V_{\alpha 2} V^{*}_{\beta 2} 
\left[ e^{ - i \frac{ ( \lambda_{2} - \lambda_{1} ) }{2E} x} - 1 \right], 
\label{ZS-amplitude-matter}
\end{eqnarray}
by which the $S$ matrix can be written, after a phase redefinition, as 
\begin{eqnarray} 
&& 
S_{\alpha \beta} 
= \delta_{\alpha \beta} + S_{\alpha \beta}^{A} + S_{\alpha \beta}^{S}. 
\end{eqnarray}
Then, setting and addressing the problem of amplitude interference can be done in a way exactly in parallel with the way we did in vacuum. 

Since the total Hamiltonian of the system is diagonalized by the $V$ matrix with the eigenvalues $\lambda_{i}$, one can argue that the decomposition \eqref{ZS-amplitude-matter} {\em is} the correct general solution to the amplitude decomposition in an arbitrary constant matter potential. The only problem for us is the lack of clear physical interpretation of the ``A'' and ``S'' variables over the entire kinematical phase space. We will revisit this point in section~\ref{sec:DMP}.

\subsection{How do we treat the ZS-type decomposition formula?}
\label{sec:what-needed}

The key issue with the ZS-type construction for us is, therefore, how and in which circumstances one can interpret the decomposed amplitudes as the matter-dressed atmospheric and the solar amplitudes. Since we are taking the bottom-up approach, in the rest of this paper, we use the amplitude decomposition formula eq.~\eqref{ZS-amplitude-matter} as a guide to proceed. That is, we impose the kinematical structure of the atmospheric and solar amplitudes in eq.~\eqref{ZS-amplitude-matter} when we carry out the amplitude decomposition. It is how we have a successful decomposition at around the solar-scale enhanced oscillations in the next section. In section~\ref{sec:DMP}, we fully utilize the ZS-type decomposition to construct the amplitude decomposition in more generic environment. 

In this context we would like to recapitulate our remark at the end of section~\ref{sec:amplitude-decomposition-AKS} that the factor $( - i \Delta_{21} x)$ in the solar amplitude in the first-order AKS perturbation theory can be understood as $\left( e^{ - i \Delta_{21} x } -1 \right)$ in an excellent approximation. Therefore, the ZS structure, in fact, had already been anticipated by the AKS amplitude decomposition. 

What happens if we treat the case of infinitesimal matter potential discussed in section~\ref{sec:decomposition-small-matter}? The problematic term, the third term in eq.~\eqref{Semu-matter-P-1st}, can be decomposed into the atmospheric and the solar amplitudes in the way the ZS decomposition dictates. Notice that the diagrammatic understanding of this term shown in eq.~\eqref{diagram-structure} involves transitions in the $2-3$ subspace, and therefore the dynamics involved is the atmospheric transition in nature. But, it contains the solar oscillation component due to involvement of the $\left( e^{ - i \Delta_{21} x } -1 \right)$ wave. 

\section{Amplitude decomposition in the solar-resonance perturbation theory}
\label{sec:solar-resonance-P}

We believe it worthwhile to explore now the region of the solar-scale enhanced oscillations in the context of amplitude decomposition in matter \cite{Martinez-Soler:2020}. For this purpose we use the ``solar-resonance perturbation theory'' formulated in ref.~\cite{Martinez-Soler:2019nhb}. It is a perturbative framework valid in region around the solar-scale oscillations where $\Delta_{21} x \sim \mathcal{O} (1)$ and 
\begin{eqnarray} 
r_{a}^{ \text{sol} } \equiv \frac{a}{\Delta m^{2}_{21}} = \frac{ \Delta_{a} }{ \Delta_{21} } \sim \mathcal{O} (1).
\label{ra-sol-def}
\end{eqnarray}
The framework has an effective expansion parameter 
\begin{eqnarray}
A_{ \text{exp} } 
&\equiv& 
c_{13} s_{13} 
\biggl | \frac{ a }{ \Delta m^2_{31} } \biggr | %%expansion-parameter
%\nonumber \\&=&
%0.8875 \times 10^{-2} x 0.146 = 0.1296 \times 10^{-2} x (3/2.8) 2 (200MeV) = 0.278 \times 10^{-2} 
= 2.78 \times 10^{-3} 
\left(\frac{ \Delta m^2_{31} }{ 2.4 \times 10^{-3}~\mbox{eV}^2}\right)^{-1}
\left(\frac{\rho}{3.0 \,\text{g/cm}^3}\right) \left(\frac{E}{200~\mbox{MeV}}\right), 
\nonumber \\
\label{expansion-parameter}
\end{eqnarray}
which guarantees smallness of the perturbative corrections, as confirmed in ref.~\cite{Martinez-Soler:2019nhb}. Since the formulation of the solar-resonance perturbation theory is done in a step-by-step manner in ref.~\cite{Martinez-Soler:2019nhb} we can just utilize here the formulas derived in that reference. 

\subsection{Amplitude decomposition in the solar oscillation region}

Since we are interested in the conceptual issue in this paper, the leading-order expression is sufficient. For more detailed properties of the decomposition see ref.~\cite{Martinez-Soler:2020}. 
The zeroth order flavor basis $S$ matrix element $S_{e \mu}^{(0)}$ is given by \cite{Martinez-Soler:2019nhb}
\begin{eqnarray} 
\hspace{-6mm}
S_{e \mu}^{(0)}
&=&
c_{23} c_{13} c_{\varphi} s_{\varphi} \left( e^{ - i h_{2} x } - e^{ - i h_{1} x } \right) 
- s_{23} e^{ - i \delta} c_{13} s_{13}  
\left( c^2_{\varphi} e^{ - i h_{1} x } + s^2_{\varphi} e^{ - i h_{2} x } - e^{ - i h_{3} x } \right).
\label{Semu-0th-solar-res} 
\end{eqnarray}
The eigenvalues of the Hamiltonian are obtained as 
\begin{eqnarray} 
h_{1} &=& 
\frac{ \Delta_{21} }{ 2 } 
\left[
\left( 1 + c^2_{13} r_{a}^{ \text{sol} } \right) 
- \sqrt{ \left( \cos 2\theta_{12} - c^2_{13} r_{a}^{ \text{sol} } \right)^2 +  \sin^2 2\theta_{12} } 
\right],
\nonumber \\
h_{2} &=&
\frac{ \Delta_{21} }{ 2 } 
\left[
\left( 1 + c^2_{13} r_{a}^{ \text{sol} } \right) 
+ \sqrt{ \left( \cos 2\theta_{12} - c^2_{13} r_{a}^{ \text{sol} } \right)^2 +  \sin^2 2\theta_{12} } 
\right], 
\nonumber \\
h_{3} &=& 
\Delta_{31} + s^2_{13} \Delta_{a}, 
\label{eigenvalues-solar-res}
\end{eqnarray}
where $r_{a}^{ \text{sol} }$ is defined in \eqref{ra-sol-def}. The angle $\varphi$ is nothing but $\theta_{12}$ in matter.

Now, let us decompose the $S_{e \mu}^{(0)}$ in eq.~\eqref{Semu-0th-solar-res} into the atmospheric and the solar amplitudes. The dominant term is the solar-scale oscillation and the atmospheric oscillation is a perturbation. But, unlike the case of the AKS expansion, we have $h_{3} \gg h_{1} \sim h_{2}$, which implies that the characteristic frequency of the perturbation is much larger, not smaller, than that of the dominant term. Then, we need a new way of isolating the solar amplitude. It is natural to take the limit $h_{3} x \rightarrow \infty$, or $\Delta_{31} x \rightarrow \infty$, which sends the atmospheric degrees of freedom high enough in energy, letting it decouple from the system. 
Since we work in the region $h_{1} x \sim \Delta_{21} x \sim \mathcal{O} (1)$, the limit implies $\Delta_{31} / \Delta_{21} \sim \Delta_{31} / a \rightarrow \infty$, or $a / \Delta_{31} \rightarrow 0$, keeping $r_{a}^{ \text{sol} }$ finite. Assuming the finite energy (and spatial) resolution $e^{ - i h_{3} x } \sim 0$ in this limit: Fast oscillations are averaged out. In this case the solar amplitude is given by eq.~\eqref{Semu-0th-solar-res}, apart from omitting the last $e^{ - i h_{3} x }$ term. 

However, the thereby obtained result of $S_{e \mu}^{ \text{sol} }$ is unsatisfactory for a number of reasons. It does not vanish at $x=0$, which means that $S_{e \mu}^{ \text{sol} }$ obtained in this way cannot be regarded as the physical oscillation amplitude. Furthermore, we find that taking the vacuum limit does not reproduce the result given in section~\ref{sec:decomposition-vacuum}.

Now, we appeal to the ZS-type construction of amplitude decomposition. It dictates that we must decompose the $S$ matrix in terms of two amplitudes, 
\begin{eqnarray} 
\left[ e^{ - i ( h_{3} - h_{1} ) x } - 1 \right], 
\hspace{6mm} 
\text{and}
\hspace{6mm}
\left[ e^{ - i ( h_{2} - h_{1} ) x } - 1 \right]. 
\end{eqnarray}
If we follow this prescription the decomposed amplitudes with the re-phasing removing $e^{ - i h_{1} x }$ read \cite{Martinez-Soler:2020} 
\begin{eqnarray} 
\left( S^{ \text{atm} }_{e \mu} \right)^{(0)}
&=&
s_{23} e^{ - i \delta} c_{13} s_{13} 
\left( e^{ - i ( h_{3} - h_{1} ) x } - 1 \right), 
\nonumber \\ 
\left( S^{ \text{sol} }_{e \mu} \right)^{(0)}
&=& 
c_{13} s_{\varphi} 
\left( c_{23} c_{\varphi} - s_{23} s_{\varphi} s_{13} e^{ - i \delta} \right) 
\left( e^{ - i ( h_{2} - h_{1} ) x } - 1 \right).
\label{Semu-atm-sol-0th-solar-res}
\end{eqnarray}
It conserves the spirit of our above discussion, but the ``kinematical'' structure that must be possessed by the decomposed amplitudes are also maintained. It nicely reproduces the vacuum result given in section~\ref{sec:decomposition-vacuum}. 

Thus, apparently the ZS-type decomposition of the $S$ matrix into the atmospheric and solar amplitudes works. Remember that we remain in the kinematic region where we know how the atmospheric and the solar oscillations are modified by the matter effect, and therefore we can rely on the ZS-type  decomposition with no reservation. 

\subsection{Non-interference and interference terms in the zeroth-order probability }
\label{sec:probability-solar-res-P}

The non-interference and interference terms in the zeroth-order oscillation probability read 
\begin{eqnarray} 
&& 
P(\nu_\mu \rightarrow \nu_e)^{ \text{non-int-fer} } 
= \biggl | \left( S_{e \mu}^{ \text{atm} } \right)^{(0)} \biggr |^2 
+ \biggl | \left( S_{e \mu}^{ \text{sol} } \right)^{(0)} \biggr |^2
\nonumber \\ 
&=& 
\left[ c^2_{13} c^2_{23} \sin^2 2 \varphi  
+ s^2_{23} \sin^2 2\theta_{13} s^4_{\varphi} 
- 8 s^2_{\varphi} J_{mr} \cos \delta \right] 
\sin^2 \frac{ ( h_{2} - h_{1} ) x }{2} 
\nonumber \\ 
&+&
4 s^2_{23} c^2_{13} s^2_{13} \sin^2 \frac{ ( h_{3} - h_{1} ) x }{2}, 
\nonumber
\end{eqnarray}
\begin{eqnarray} 
&& 
P(\nu_\mu \rightarrow \nu_e)^{ \text{int-fer} } 
= 2 \mbox{Re} 
\biggl[ 
\left\{ \left( S_{e \mu}^{ \text{atm} } \right)^{(0)} \right\}^* 
\left( S_{e \mu}^{ \text{sol} } \right)^{(0)} \biggr] 
\nonumber \\ 
&=&
\left( 4 J_{mr} \cos \delta - s^2_{23} \sin^2 2\theta_{13} 
s^2_{\varphi} \right) 
\left[
- \sin^2 \frac{ ( h_{3} - h_{2} ) x }{2} + \sin^2 \frac{ ( h_{3} - h_{1} ) x }{2} + \sin^2 \frac{ ( h_{2} - h_{1} ) x }{2} \right]
\nonumber \\ 
&-& 
8 J_{mr} \sin \delta 
\sin \frac{ ( h_{3} - h_{1} ) x }{2} \sin \frac{ ( h_{2} - h_{1} ) x }{2} 
\sin \frac{ ( h_{3} - h_{2} ) x }{2},
\label{P-mue-decomposition-solar-R}
\end{eqnarray}
where we have defined the ``matter-dressed Jarlskog'' factor 
\begin{eqnarray} 
J_{mr} \equiv c_{23} s_{23} c^2_{13} s_{13} c_{\varphi} s_{\varphi}.
\label{matter-Jarlskog-1st}
\end{eqnarray}
It appears that the $\delta$-dependent terms are dominant in the solar oscillation region as well. 

\subsection{$\varphi$ symmetry as a quantum mechanics protecting symmetry}
\label{sec:QM-protect} 

We have noticed that the oscillation probability possesses the $\varphi$ symmetry, an invariance under the transformation $\varphi \rightarrow \varphi + \frac{\pi}{2}$ \cite{Martinez-Soler:2019nhb}. See also ref.~\cite{Denton:2016wmg}. The nature of the $\varphi$ symmetry is identified as the ``dynamical'' symmetry, not the symmetry of the Hamiltonian \cite{Martinez-Soler:2019nhb}. Or, in other word, it is a reparametrization invariance of the variable that is born out of the construction of perturbation theory. 

Notice that each one of the decomposed probabilities in eq.~\eqref{P-mue-decomposition-solar-R} violates the $\varphi$ symmetry, which existed  in the total probability, $P(\nu_\mu \rightarrow \nu_e) = P(\nu_\mu \rightarrow \nu_e)^{ \text{non-int-fer} } + P(\nu_\mu \rightarrow \nu_e)^{ \text{int-fer} }$, the quantum mechanical observable. Therefore, if we enforce the $\varphi$ symmetry, quantum mechanics, i.e., $q=1$ is the unique choice that is allowed.

Namely, the $\varphi$ symmetry ``protects'' the size of the interference term to be the one dictated by quantum mechanics. 

In section~\ref{sec:analysis-method}, we have described the way of analyzing data to test at what significance the case of no interference is disfavored by creating the test probability $P(\nu_\beta \rightarrow \nu_\alpha: q) = P(\nu_\beta \rightarrow \nu_\alpha)^{ \text{non-int-fer} } + q P(\nu_\beta \rightarrow \nu_\alpha)^{ \text{int-fer} }$. Thereby defined test probability violates the $\varphi$ symmetry for $q \neq 1$. Nonetheless, we still believe that the analysis procedure is tenable because the $\varphi$ symmetry is the dynamical symmetry, not the symmetry of the Hamiltonian.\footnote{
%%%%%%%%%%%%% footnote %%%%%%%%%%%%%
The term ``protects'' may be too strong if the assumed minimum at $q=1$ is shallow. On the contrary, if the minimum is very deep, we would observe a steep parabola of $\chi^2 (q)$ centered around $q=1$. In this case, the analysis procedure described in section~\ref{sec:analysis-method} becomes superfluous, as the situation of $q \neq 1$ is essentially prohibited by the symmetry, which we predict not to be the case. 
}

\section{Amplitude decomposition in the helio perturbation theory}
\label{sec:helio-perturbation-theory} 

Up to now we have had the two cases of successful amplitude decomposition based on 
the perturbative frameworks, the AKS and the solar-resonance perturbation theory. While the latter covers the region of the solar-scale enhanced oscillation, the former serves for the short- or medium-baseline accelerator neutrino experiments \cite{Abe:2019vii,Abe:2018uyc,Baussan:2013zcy}. Yet, the amplitude decomposition formula usable in region of the atmospheric-scale matter enhanced oscillation is still missing. In fact, there exist the ongoing and the upcoming LBL experiments which utilize the longer baselines, and hence have stronger matter effects due to the higher beam energies. They include MINOS/MINOS+~\cite{Adamson:2020ypy}, NO$\nu$A \cite{Acero:2019ksn}, DUNE \cite{Abi:2020evt}, and T2KK\footnote{
%%%%%%%%%%%%% footnote %%%%%%%%%%%%%
A possible acronym used in ref.~\cite{Kajita:2006bt}, but now for the updated name for the setting, ``Tokai-to-Kamioka observatory-Korea neutrino observatory''. 
} 
\cite{Abe:2016ero}. 

In this section, we discuss amplitude decomposition in the framework appropriate for application to these LBL experimental settings. The helio-to-terrestrial ratio 
\begin{eqnarray} 
\epsilon \equiv \frac{ \Delta m^2_{21} }{ \Delta m^2_{31} }, 
\label{epsilon-def}
\end{eqnarray}
which was introduced in section~\ref{sec:AKS}, now plays the role of the unique expansion parameter, and hence it may be called as the ``helio perturbation theory''. It allows us to treat the sizable matter effect as strong as the vacuum effect in a non-perturbative fashion to all orders. This is another example that the features of matter-dressed atmospheric and the solar oscillations are well understood. 
The framework first appeared in the early work \cite{Arafune:1996bt}, which is followed by the systematic exploration in refs.~\cite{Cervera:2000kp,Freund:2001pn,Akhmedov:2004ny} and is refined in \cite{Minakata:2015gra}. In this paper, we restrict ourselves to first order in the helio correction, which may be sufficient for the ongoing and the next generation LBL experiments quoted above.

To make the route to the physics discussion shorter, we defer our brief recollection of the formulation of the helio perturbation theory into appendix~\ref{sec:formulation-helio-perturbation}.

\subsection{Amplitude decomposition in the $\nu_{\mu} \rightarrow \nu_{e}$ channel}  
\label{sec:decomposition-helio-P-mue} 

In the helio perturbation theory, the zeroth- and the first-order amplitudes in the $\nu_{\mu} \rightarrow \nu_{e}$ channel are given by 
\begin{eqnarray} 
S_{e \mu}^{(0)} 
&=& 
s_{23} c_{\phi} s_{\phi} e^{ - i \delta} \left( e^{ - i h_{3} x } - e^{ - i h_{1} x } \right),  
\nonumber \\ 
S_{e \mu}^{(1)} &=& 
s^2_{12} s_{23} e^{ - i \delta} c_{\phi} s_{\phi} 
( -i \Delta_{21} x ) 
\biggl[
s^2_{13} \left( e^{ - i h_{3} x } - e^{ - i h_{1} x } \right)
\nonumber \\ 
&+&
\cos 2\theta_{13} 
\left( s^2_{\phi} e^{ - i h_{3} x } - c^2_{\phi} e^{ - i h_{1} x } \right) 
- c_{\phi} s_{\phi} \sin 2\theta_{13}  
\left( e^{ - i h_{3} x } + e^{ - i h_{1} x } \right) 
\biggr] 
\nonumber \\ 
&+&
c_{23} c_{12} s_{12} s_{\phi} \sin (\phi - \theta_{13} )
\left( \frac{ \Delta_{21} }{ h_{3} -  h_{2} } \right) 
\left( e^{ - i h_{3} x } - e^{ - i h_{2} x } \right)
\nonumber \\ 
&+& 
c_{23} c_{12} s_{12} c_{\phi} \cos (\phi - \theta_{13} )
\left( \frac{ \Delta_{21} }{ h_{2} -  h_{1} } \right) 
\left( e^{ - i h_{2} x } - e^{ - i h_{1} x } \right) 
\nonumber \\
&+& 
s^2_{12} s_{23} e^{ - i \delta} 
\left[ - c_{13} s_{13} 
+ c_{\phi} s_{\phi} 
\cos 2 (\phi - \theta_{13}) 
\right] 
\left( \frac{ \Delta_{21} }{ h_{3} -  h_{1} } \right) 
\left( e^{ - i h_{3} x } - e^{ - i h_{1} x } \right).
\label{Smatrix-0th-1st-helio-P}
\end{eqnarray}
In eq.~\eqref{Smatrix-0th-1st-helio-P}, $\phi$ denotes $\theta_{13}$ in matter and is defined in eq.~\eqref{cos-sin-2phi}. $h_{i}$ ($i=1,2,3$) are the eigenvalues of the zeroth-order Hamiltonian eq.~\eqref{tilde-H-F-def}. The formulation and some computational details are given in appendix~\ref{sec:formulation-helio-perturbation}. 

To implement the ZS structure we factor out $e^{ - i h_{1} x }$ from the oscillation $S$ matrix as $S = e^{ - i h_{1} x } \tilde{S}$ and rename $\tilde{S}$ as the new $S$ matrix. Then, the decomposition to the atmospheric and the solar amplitudes can be performed to give the following expressions: 
In zeroth-order $\left( S_{e \mu}^{ \text{sol} } \right)^{(0)} = 0$ and 
\begin{eqnarray} 
\left( S_{e \mu}^{ \text{atm} } \right)^{(0)}
&=& 
s_{23} c_{\phi} s_{\phi} e^{ - i \delta} 
\left( e^{ - i ( h_{3} - h_{1} ) x } - 1 \right). 
\label{amp-decompose-helio-P-mue-0th}
\end{eqnarray}
In first order the decomposition reads: 
\begin{eqnarray} 
\left( S_{e \mu}^{ \text{atm} } \right)^{(1)} 
&=&
\biggl\{
s^2_{12} s_{23} e^{ - i \delta} c_{\phi} s_{\phi} \sin^2 ( \phi - \theta_{13} ) 
( -i \Delta_{21} x ) 
+ c_{23} c_{12} s_{12} s_{\phi} \sin (\phi - \theta_{13} )
\left( \frac{ \Delta_{21} }{ h_{3} -  h_{2} } \right) 
\nonumber \\
&+& 
s^2_{12} s_{23} e^{ - i \delta} 
\left[ - c_{13} s_{13} 
+ c_{\phi} s_{\phi} 
\cos 2 (\phi - \theta_{13}) 
\right] 
\left( \frac{ \Delta_{21} }{ h_{3} -  h_{1} } \right) 
\biggr\}
\left( e^{ - i ( h_{3} - h_{1} ) x } - 1 \right), 
\nonumber \\ 
\left( S_{e \mu}^{ \text{sol} } \right)^{(1)} 
&=&
- s^2_{12} s_{23} e^{ - i \delta} c_{\phi} s_{\phi} \cos 2 ( \phi - \theta_{13} ) 
( -i \Delta_{21} x ) 
\nonumber \\ 
&-& 
c_{23} c_{12} s_{12} 
\biggl\{
s_{\phi} \sin (\phi - \theta_{13} )
\left( \frac{ \Delta_{21} }{ h_{3} -  h_{2} } \right) 
- c_{\phi} \cos (\phi - \theta_{13} )
\left( \frac{ \Delta_{21} }{ h_{2} -  h_{1} } \right) 
\biggr\}
\left( e^{ - i ( h_{2} - h_{1} ) x } - 1 \right). 
\nonumber \\ 
\label{amp-decompose-helio-P-mue-1st}
\end{eqnarray}
Notice that the $( -i \Delta_{21} x )$ term is naturally in the solar amplitude because one can interpret it as $( -i \Delta_{21} x ) \approx \left( e^{ - i \Delta_{21} x } -1 \right)$, as mentioned in sections~\ref{sec:amplitude-decomposition-AKS} and~\ref{sec:principle} in the context of the AKS perturbation theory. 

\subsection{Amplitude decomposition in the $\nu_{\mu} \rightarrow \nu_{\tau}$ channel}  
\label{sec:decomposition-helio-P-mu-tau}

Similarly, the amplitude decomposition in the $\nu_{\mu} \rightarrow \nu_{\tau}$ channel in the zeroth and first order is given by 
\begin{eqnarray} 
&& 
\left( S_{\tau \mu}^{ \text{atm} } \right)^{(0)} 
= c_{23} s_{23} c^2_{\phi} \left( e^{ - i ( h_{3} - h_{1} ) x } - 1 \right), 
\nonumber \\
&&
\left( S_{\tau \mu}^{ \text{sol} } \right)^{(0)} 
= - c_{23} s_{23} \left( e^{ - i ( h_{2} - h_{1} ) x } - 1 \right),
\label{amp-decompose-helio-P-mutau-0th}
\end{eqnarray}
and 
\begin{eqnarray} 
\left( S_{\tau \mu}^{ \text{atm} } \right)^{(1)} 
&=& 
\biggl[
c_{23} s_{23} s^2_{12} c^2_{\phi} \sin^2 ( \phi - \theta_{13} ) 
( - i \Delta_{21} x ) 
- c_{23} s_{23} s^2_{12} c_{\phi} s_{\phi} 
\sin 2 ( \phi - \theta_{13} ) 
\frac{ \Delta_{21} }{ ( h_{3} -  h_{1} ) } 
\nonumber \\
&+& 
c_{12} s_{12} c_{\phi} \sin ( \phi - \theta_{13} ) 
\left( \cos 2\theta_{23} \cos \delta + i \sin \delta \right)
\frac{ \Delta_{21} }{ ( h_{3} -  h_{2} ) } 
\biggr]
\left( e^{ - i ( h_{3} - h_{1} ) x } - 1 \right), 
\nonumber 
\end{eqnarray}
\begin{eqnarray} 
\left( S_{\tau \mu}^{ \text{sol} } \right)^{(1)} 
&=&
c_{23} s_{23} 
\left[ ( s^2_{13} s^2_{12} - c^2_{12} ) + s^2_{12} c_{\phi} s_{\phi} 
\sin 2 ( \phi - \theta_{13} ) \right] ( - i \Delta_{21} x ) 
\nonumber \\ 
&-& 
\biggl[
c_{23} s_{23} c^2_{12} 
( - i \Delta_{21} x ) 
+ c_{12} s_{12} \left( \cos 2\theta_{23} \cos \delta + i \sin \delta \right) 
\nonumber \\ 
&\times& 
\biggl\{
c_{\phi} \sin ( \phi - \theta_{13} ) 
\frac{ \Delta_{21} }{ ( h_{3} -  h_{2} ) }  
+ s_{\phi} \cos ( \phi - \theta_{13} )
\frac{ \Delta_{21} }{ ( h_{2} -  h_{1} ) } 
\biggr\}
\biggr]
\left( e^{ - i ( h_{2} - h_{1} ) x } - 1 \right). 
\nonumber \\ 
\label{amp-decompose-helio-P-mutau-1st}
\end{eqnarray}
We have checked that the amplitude decomposition in the $\nu_{\mu} \rightarrow \nu_{e}$ and $\nu_{\mu} \rightarrow \nu_{\tau}$ channels, eqs.~\eqref{amp-decompose-helio-P-mue-0th} - \eqref{amp-decompose-helio-P-mutau-1st}, have the correct vacuum limit.

\subsection{Non-interference and interference terms in the probability to first order }
\label{sec:probability-helio-P}

Here we present the decomposed probabilities $P(\nu_{\mu} \rightarrow \nu_{\alpha})^{ \text {non-int-fer} }$ and $P(\nu_{\mu} \rightarrow \nu_{\alpha})^{ \text {int-fer} }$ ($\alpha = e, \tau$) and discuss the $\nu_{\mu} \rightarrow \nu_{e}$ and $\nu_{\mu} \rightarrow \nu_{\tau}$ channels in parallel for comparison. 

In the $\nu_{\mu} \rightarrow \nu_{e}$ channel they are given to first order as
\begin{eqnarray} 
&& 
P(\nu_{\mu} \rightarrow \nu_{e})^{ \text {non-int-fer} } 
=
\biggl[
s^2_{23} \sin^2 2\phi 
+ 8 \tilde{J}_{mr}^{s} \cos \delta 
\left( \frac{ \Delta_{21} }{ h_{3} -  h_{2} } \right)  
\nonumber \\ 
&+& 
2 s^2_{23} s^2_{12} \sin 2\phi 
\left\{ - \sin 2\theta_{13} + \sin 2\phi \cos 2 (\phi - \theta_{13}) \right\}
\left( \frac{ \Delta_{21} }{ h_{3} -  h_{1} } \right) 
\biggr]
\sin^2 \frac{ ( h_{3} - h_{1} ) x}{2}, 
\label{Pmue-non-int-fer-helio-P}
\end{eqnarray}
\begin{eqnarray} 
&& 
P(\nu_{\mu} \rightarrow \nu_{e})^{ \text {int-fer} } 
\nonumber \\ 
&=& 
- 2 s^2_{23} s^2_{12} c^2_{\phi} s^2_{\phi} 
\cos 2 \left( \phi - \theta_{13} \right)
( \Delta_{21} x ) 
\sin ( h_{3} - h_{1} ) x
\nonumber \\ 
&-& 
8 \biggl\{
\tilde{J}_{mr}^{s} \left( \frac{ \Delta_{21} }{ h_{3} -  h_{2} } \right)  
- \tilde{J}_{mr}^{c}  \left( \frac{ \Delta_{21} }{ h_{2} -  h_{1} } \right) 
\biggr\} 
\cos \left\{ \delta + \frac{ ( h_{3} - h_{2} ) x }{2} \right\}
\sin \frac{ ( h_{2} - h_{1} ) x }{2} \sin \frac{ ( h_{3} - h_{1} ) x }{2}.
\nonumber \\
\label{Pmue-int-fer-helio-P}
\end{eqnarray}
and in the $\nu_{\mu} \rightarrow \nu_{\tau}$ channel
\begin{eqnarray} 
&& 
P(\nu_{\mu} \rightarrow \nu_{\tau})^{ \text {non-int-fer} } 
\nonumber \\ 
&=& 
4 c^2_{23} s^2_{23} 
\left[ c^4_{\phi} \sin^2 \frac{ ( h_{3} - h_{1} ) x }{2} 
+ \sin^2 \frac{ ( h_{2} - h_{1} ) x }{2} \right]
\nonumber \\ 
&-&
2 c^2_{23} s^2_{23} 
\left[ ( s^2_{13} s^2_{12} - c^2_{12} ) + s^2_{12} c_{\phi} s_{\phi} 
\sin 2 ( \phi - \theta_{13} ) \right] 
( \Delta_{21} x ) \sin ( h_{2} - h_{1} ) x 
\nonumber \\ 
&-&
8 c^2_{23} s^2_{23} s^2_{12} c^3_{\phi} s_{\phi} 
\sin 2 ( \phi - \theta_{13} ) 
\frac{ \Delta_{21} }{ ( h_{3} -  h_{1} ) } 
\sin^2 \frac{ ( h_{3} - h_{1} ) x }{2} 
\nonumber \\
&+&
8 \tilde{J}_{mrs}^{s} \cos 2\theta_{23} \cos \delta 
\frac{ \Delta_{21} }{ ( h_{3} -  h_{2} ) }  
\left\{
c^2_{\phi} 
\sin^2 \frac{ ( h_{3} - h_{1} ) x }{2} 
+ \sin^2 \frac{ ( h_{2} - h_{1} ) x }{2} 
\right\} 
\nonumber \\
&+&
8 \tilde{J}_{mrs}^{c} \cos 2\theta_{23} \cos \delta 
\frac{ \Delta_{21} }{ ( h_{2} -  h_{1} ) } 
\sin^2 \frac{ ( h_{2} - h_{1} ) x }{2},
\label{Pmutau-non-int-fer-helio-P}
\end{eqnarray}
\begin{eqnarray} 
&& 
P(\nu_{\mu} \rightarrow \nu_{\tau})^{ \text {int-fer} } 
\nonumber \\ 
&=& 
2 c^2_{23} s^2_{23} c^2_{\phi} 
\left[ ( s^2_{13} s^2_{12} - c^2_{12} ) + s^2_{12} c_{\phi} s_{\phi} 
\sin 2 ( \phi - \theta_{13} ) \right] 
( \Delta_{21} x ) \sin ( h_{3} - h_{1} ) x 
\nonumber \\ 
&-& 
8 \biggl[
c^2_{23} s^2_{23} \left\{ 
c^2_{\phi} - s^2_{12} c_{\phi} s_{\phi} 
\sin 2 ( \phi - \theta_{13} ) 
\frac{ \Delta_{21} }{ ( h_{3} -  h_{1} ) } 
\right\} 
\nonumber \\
&+& 
\cos 2\theta_{23} \cos \delta 
\left\{
\left( 1 + c^2_{\phi} \right)
\tilde{J}_{mrs}^{s} \frac{ \Delta_{21} }{ ( h_{3} -  h_{2} ) } 
+ c^2_{\phi} \tilde{J}_{mrs}^{c} \frac{ \Delta_{21} }{ ( h_{2} -  h_{1} ) } 
\right\}
\biggr]
\nonumber \\
&\times& 
\sin \frac{ ( h_{2} - h_{1} ) x }{2} 
\sin \frac{ ( h_{3} - h_{1} ) x }{2}
\cos \frac{ ( h_{3} - h_{2} ) x }{2} 
\nonumber \\
&+& 
8 \biggl[ 
- c^2_{23} s^2_{23} c^2_{\phi} 
\left\{ c^2_{12} - s^2_{12} \sin^2 ( \phi - \theta_{13} ) \right\} 
( \Delta_{21} x ) 
\nonumber \\
&+& 
\sin \delta 
\left\{
- \tilde{J}_{mr}^{s} \frac{ \Delta_{21} }{ ( h_{3} -  h_{2} ) } 
+ \tilde{J}_{mr}^{c} \frac{ \Delta_{21} }{ ( h_{2} -  h_{1} ) } 
\right\} 
\biggr]
\sin \frac{ ( h_{3} - h_{1} ) x }{2} 
\sin \frac{ ( h_{2} - h_{1} ) x }{2} 
\sin \frac{ ( h_{3} - h_{2} ) x }{2}.
\nonumber \\
\label{Pmutau-int-fer-helio-P}
\end{eqnarray}
In the above equations we have introduced, in addition to the one in eq.~\eqref{matter-Jarlskog-1st}, the following four Jarlskog factors in matter:
\begin{eqnarray} 
&&
\tilde{J}_{mr}^{s} \equiv 
c_{23} s_{23} c_{12} s_{12} c_{\phi} s^2_{\phi} \sin (\phi - \theta_{13} ) 
= J_r 
\frac{ - 1 + r_{a} + \sqrt{ ( \cos 2\theta_{13} - r_{a} )^2 + \sin^2 2\theta_{13} } }{ 2 \left[ ( \cos 2\theta_{13} - r_{a} )^2 + \sin^2 2\theta_{13} \right] }, 
\nonumber \\
&& 
\tilde{J}_{mr}^{c} \equiv 
c_{23} s_{23} c_{12} s_{12} c^2_{\phi} s_{\phi} \cos (\phi - \theta_{13} ) 
= J_r 
\frac{ 1 - r_{a} + \sqrt{ ( \cos 2\theta_{13} - r_{a} )^2 + \sin^2 2\theta_{13} }  }{ 2 \left[ ( \cos 2\theta_{13} - r_{a} )^2 + \sin^2 2\theta_{13} \right] }, 
\label{matter-Jarlskog-2nd}
\end{eqnarray} 
\begin{eqnarray} 
&&
\tilde{J}_{mrs}^{s} \equiv 
c_{23} s_{23} c_{12} s_{12} c_{\phi} \sin (\phi - \theta_{13} ) 
= 
J_{rs} \frac{ 1 + r_{a} - \sqrt{ 1 + r_{a}^2 - 2 r_{a} \cos 2\theta_{13} }  }{ 2 \sqrt{ 1 + r_{a}^2 - 2 r_{a} \cos 2\theta_{13} } }, 
\nonumber \\
&& 
\tilde{J}_{mrs}^{c} \equiv 
c_{23} s_{23} c_{12} s_{12} s_{\phi} \cos (\phi - \theta_{13} ) 
=
J_{rs} \frac{ 1 + r_{a} + \sqrt{ 1 + r_{a}^2 - 2 r_{a} \cos 2\theta_{13} }  }{ 2 \sqrt{ 1 + r_{a}^2 - 2 r_{a} \cos 2\theta_{13} } }.
\label{matter-Jarlskog-3rd}
\end{eqnarray} 
where $r_{a} = \Delta_{a} / \Delta_{31} = a / \Delta m^2_{31}$, as defined in eq.~\eqref{matter-vacuum-ratio}. 
The ``matter-Jarlskog'' factors in eqs.~\eqref{matter-Jarlskog-2nd} and \eqref{matter-Jarlskog-3rd} are the matter-dressed versions of $J_r = c_{23} s_{23} c_{12} s_{12} c^2_{13} s_{13}$ and the ``$c^2_{13}$-missed'' one $J_{rs} = c_{23} s_{23} c_{12} s_{12} s_{13}$, respectively, which are defined in eq.~\eqref{Jr-def}. As is well known, the $\sin \delta$ terms must be proportional to $J_{r}$, as dictated by the Naumov identity~\cite{Naumov:1991ju}. $J_{rs}$ appears in the probabilities in vacuum in the $\nu_{\mu} - \nu_{\tau}$ sector, as seen in section~\ref{sec:decomposition-vacuum}. In fact, one can show generally that the $\cos \delta$ terms must be proportional to $J_{rs}$~\cite{Asano:2011nj} in all the oscillation channels.\footnote{
%%%%%%%%%%%%% 
In the $\nu_{\mu} \rightarrow \nu_{e}$ channel, it is empirically known that the $J_{rs}$ dependence of the $\cos \delta$ terms is elevated to the $J_{r}$ dependence.
} 
The explicit forms given in the right-hand sides of eqs.~\eqref{matter-Jarlskog-2nd} and \eqref{matter-Jarlskog-3rd} guarantee that these general features hold.

Again, $\phi \rightarrow \phi + \frac{\pi}{2}$ symmetry \cite{Martinez-Soler:2018lcy,Martinez-Soler:2019nhb}, which exists in the total probability, is broken when the probability is decomposed into the non-interference and interference parts. That is, the $\phi$ symmetry is also the ``protecting symmetry'' for quantum mechanical interference. Since a little complicated reduction is needed to show the $\phi$ invariance of the total probability $P(\nu_{\mu} \rightarrow \nu_{\tau}) = P(\nu_{\mu} \rightarrow \nu_{\tau})^{ \text {non-int-fer} } + P(\nu_{\mu} \rightarrow \nu_{\tau})^{ \text {int-fer} }$, we present its explicit form in appendix~\ref{sec:P-mutau-matter}.

As in vacuum the dominant effects in $P(\nu_{\mu} \rightarrow \nu_{e})^{ \text {int-fer} }$ is from the $\delta$ dependent terms as the first term in eq.~\eqref{Pmue-int-fer-helio-P} is suppressed by $s^2_{\phi} \simeq s^2_{13}$. Similarly, the dominant term in $P(\nu_{\mu} \rightarrow \nu_{\tau})^{ \text {int-fer} }$ is the $\delta$-independent term. The feature is the same in $P(\nu_{\mu} \rightarrow \nu_{\mu})^{ \text {int-fer} }$, though not shown in this paper. These features are akin to those possessed by the interference terms in vacuum, as seen in section~\ref{sec:decomposition-vacuum}. 

\section{Amplitude decomposition in more generic environment: From ZS to DMP construction}
\label{sec:DMP}

We have stressed in section~\ref{sec:principle} that identifying the relevant dynamical variables in a given kinematical space and clarifying their physical meaning are highly nontrivial issues. To our knowledge, the general and physically appealing answer to this question does not appear to be known. Independently of the ZS approach introduced in section~\ref{sec:principle}, there exists an attempt by Akhmedov, Maltoni, and Smirnov (AMS) to identify the physically motivated A and S variables~\cite{Akhmedov:2008qt}. They calculated the decomposed amplitudes $S_{\alpha \beta}^{A}$ and $S_{\alpha \beta}^{S}$ as a function of the matter-dressed atmospheric and solar variables under the uniform matter-density approximation, and discussed physics of the interference in the context of atmospheric neutrino experiments. 

In this paper we have taken a ``bottom-up'' approach to the amplitude decomposition. After examining various perturbative schemes whose regions of validity span the solar- or the atmospheric-scale enhancements, we have naturally arrived at our own proposal for the solution to the problem of amplitude decomposition in more generic environment. Here, ``generic environment'' means either the region of energies and matter densities in which the $A$ and $S$ variables can be interpreted as those of the matter-dressed atmospheric and solar oscillation modes, or the ones outside of it. 

By following the Jacobi method first introduced to describe neutrino oscillations in ref.~\cite{Agarwalla:2013tza}, Denton {\it et al.}~\cite{Denton:2016wmg} formulated a framework in which the eigenvalues and the $V$ matrix elements can be expressed by the two matter-dressed mixing angles $\theta_{13}$ and $\theta_{12}$, and the matter-undressed $\theta_{23}$ and $\delta$. We call the amplitude decomposition scheme based on the Denton {\it et al.} framework as the DMP decomposition. Following ref.~\cite{Denton:2016wmg}, in this section we take the ATM convention of the mixing matrix $U_{\text{\tiny MNS}}$ in which $e^{ \pm i \delta}$ is attached to $s_{23}$. 

\subsection{Amplitude decomposition based on the DMP framework}
\label{sec:DMP-decomposition}

We define the DMP amplitude decomposition by doing replacements in the eigenvalues and the $V$ matrix in the ZS amplitudes in \eqref{ZS-amplitude-matter}:\footnote{
%%%%%%%%%%%% footnote %%%%%%%%%%%%%%
See, however, a comment on the different phase convention below.
}
\begin{eqnarray} 
&& 
\lambda_{i} \rightarrow \lambda_{i}^{\text{\tiny DMP}}, 
\hspace{10mm} 
V \rightarrow V_{\text{\tiny DMP}}.
\label{replacements}
\end{eqnarray}
We note that the $V$ matrix method \cite{Minakata:1998bf}, which has been adopted in refs.~\cite{Minakata:2015gra,Denton:2016wmg}, makes the DMP formulation of the amplitude decomposition particularly simple. In fact, all the necessary ingredients are already computed in ref.~\cite{Denton:2016wmg} to second order in perturbation. The leading and the first-order expressions of the $V$ matrix, $V = V_{\text{\tiny DMP}}^{(0)} + V_{\text{\tiny DMP}}^{(1)}$, are given by 
\begin{eqnarray}
&& 
V_{\text{\tiny DMP}}^{(0)}
= \left[
\begin{array}{ccc}
c_{\psi} c_{\phi} & s_{\psi} c_{\phi} & s_{\phi} \\
- c_{23} s_{\psi} - s_{23} c_{\psi} s_{\phi} e^{i\delta} &
c_{23} c_{\psi} - s_{23} s_{\psi} s_{\phi} e^{i\delta} & s_{23} c_{\phi} e^{i\delta} \\
s_{23} s_{\psi} e^{- i \delta} - c_{23} c_{\psi} s_{\phi} &
- s_{23} c_{\psi} e^{- i \delta} - c_{23} s_{\psi} s_{\phi} & c_{23} c_{\phi} \\
\end{array}
\right], 
\nonumber \\
&& 
V_{\text{\tiny DMP}}^{(1)}
= V_{\text{\tiny DMP}}^{(0)} W_{1},
\label{DMP-V-matrix}
\end{eqnarray}
where $\phi$ and $\psi$ are the matter-dressed mixing angles $\theta_{13}$ and $\theta_{12}$, respectively, and $W_{1}$ is defined by
\begin{eqnarray}
&& 
W_{1} = 
\epsilon^{\prime} c_{12} s_{12} \sin ( \phi - \theta_{13} )
\left[
\begin{array}{ccc}
0 & 0 & - s_\psi 
\frac{ \Delta m^2_{ \text{ren} } }{ \lambda_{3} - \lambda_{1} } \\
0 & 0 & c_\psi 
\frac{ \Delta m^2_{ \text{ren} } }{ \lambda_{3} - \lambda_{2} } \\
s_\psi 
\frac{ \Delta m^2_{ \text{ren} } }{ \lambda_{3} - \lambda_{1} } & 
- c_\psi 
\frac{ \Delta m^2_{ \text{ren} } }{ \lambda_{3} - \lambda_{2} } & 0 \\
\end{array}
\right].
\label{W1-def}
\end{eqnarray}
In \eqref{W1-def}, $\epsilon^{\prime}$ is defined as $\epsilon^{\prime} \equiv \Delta m^2_{31} / \Delta m^2_{ \text{ren} }$ where the renormalized atmospheric $\Delta m^2$ is defined as $\Delta m^2_{ \text{ren} } \equiv \Delta m^2_{31} - s^2_{12} \Delta m^2_{21}$~\cite{Minakata:2015gra}.
$\lambda_{i}$ ($i=1,2,3$) denote the eigenvalues of $2E$ times the zeroth-order Hamiltonian, and the explicit forms of them as well as those of $\phi$ and $\psi$ are given in ref.~\cite{Denton:2016wmg}.

The necessary ingredients for constructing the DMP decomposition are completely specified by the information above to first order in perturbation. While we do not present the explicit forms of the decomposed probabilities, $P(\nu_{\beta} \rightarrow \nu_{\alpha})^{ \text {non-int-fer} }$ and $P(\nu_{\beta} \rightarrow \nu_{\alpha})^{ \text {int-fer} }$, they can be obtained by the replacements $\theta_{13} \rightarrow \phi$ and $\theta_{12} \rightarrow \psi$ in the vacuum expressions in the leading order. The prescription for computing the first-order corrections is also given above. 

Finally, we make some remarks on the following two relevant issues:  
\begin{itemize}
\item
Ambiguity in the ``atmospheric'' oscillation frequency,

\item 
Physical interpretation of the ``A'' and ``S'' variables and the region of validity of the DMP decomposition.

\end{itemize}
\noindent
In vacuum, there is a problem of how to define the effective atmospheric $\Delta m^2$. It could be $\Delta m^2_{31}$, or $\Delta m^2_{32}$, or an interpolated value in between. In the analysis to identify the interference effect in JUNO reactor neutrino experiment, we have examined the both cases of $\Delta m^2_{31}$ and $\Delta m^2_{32}$ for the atmospheric $\Delta m^2$ and obtained the same result \cite{Huber:2019frh}.\footnote{
%%%%%%%%%%%%%% 
In fact, it was confirmed during the work described in ref.~\cite{Huber:2019frh} that the values of $\chi^2 (q=0)$ are stable over varying choices in the region $0 \leq r \leq 1$ of the atmospheric $\Delta m^2 (r) = (1-r) \Delta m^2_{31} + r \Delta m^2_{32}$.
}

In matter the situation is different. The natural choice for the atmospheric frequency is determined by the system itself. We use the state label with the eigenvalues $\lambda_{1} < \lambda_{2} < \lambda_{3}$ in the normal mass ordering (NMO), and $\lambda_{3} < \lambda_{1} < \lambda_{2}$ in the inverted mass ordering (IMO). (See e.g., Fig.~1 in ref.~\cite{Denton:2016wmg}.) Since the atmospheric resonance is in the 2-3 level crossing in the NMO, it is natural to design the amplitude decomposition with rephasing factor $e^{ - i \frac{ \lambda_{2} }{2E} x}$ so that the decomposed amplitude read
\begin{eqnarray} 
&& 
S_{\alpha \beta}^{ \text{atm} } 
\equiv  
V_{\alpha 3} V^{*}_{\beta 3} 
\left[ e^{ - i \frac{ ( \lambda_{3} - \lambda_{2} ) }{2E} x} - 1 \right], 
\nonumber \\
&& 
S_{\alpha \beta}^{ \text{sol} } \equiv  
V_{\alpha 1} V^{*}_{\beta 1} 
\left[ e^{ i \frac{ ( \lambda_{2} - \lambda_{1} ) }{2E} x} - 1 \right].
\label{DMP-decompose-2}
\end{eqnarray}
In the IMO, however, the resonance is in the 1-3 level crossing, and therefore it is natural to use $e^{ - i \frac{ \lambda_{1} }{2E} x}$ rephasing as in eq.~\eqref{ZS-amplitude-matter}. Thus, there is a physics motivated way of determining the atmospheric oscillation frequency in matter. We note that, of course, the both ways of decomposition, eqs.~\eqref{ZS-amplitude-matter} and \eqref{DMP-decompose-2}, lead to the same probability, as they differ only in the overall phase. But, due to the difference in the decomposed amplitudes, the decomposed probabilities are different between the decompositions~\eqref{ZS-amplitude-matter} and \eqref{DMP-decompose-2}.

The region of validity and the physical interpretation of the DMP decomposition are the remaining important problem. Ideally, we could precisely define the kinematical phase space boundary within which the ``A'' and ``S'' variables can be interpreted as the matter-dressed atmospheric and solar oscillation variables. However, it does not appear to be possible at this moment to our understanding.\footnote{
%%%%%%%%%%%%% footnote %%%%%%%%%%%%%%
Even the definitions, what are the matter-effect modified  ``atmospheric'' and ''solar'' oscillation variables, are not obvious to the author.
}

At least, we have to check that our decomposition formula is consistent with the ones derived in the regions of atmospheric-scale and the solar-scale enhanced oscillations. Since the DMP framework generalizes the one of ref.~\cite{Minakata:2015gra} by doing another 1-2 space rotation it is very likely that it is smoothly connected to the helio perturbation theory discussed in section~\ref{sec:helio-perturbation-theory}. What is more nontrivial is the smooth connection to the solar oscillation region. However, our preliminary study shows that the decomposed amplitudes in the solar-resonance perturbation theory can be recovered by taking the appropriate limit in the DMP decomposition formulas. Therefore, it is very likely that the DMP decomposition successfully interpolates the two regions of the atmospheric- and solar-scale enhanced oscillations. In this case, the decomposition formulas, eqs.~\eqref{DMP-decompose-2} and/or \eqref{ZS-amplitude-matter}, may apply to the whole region sandwiched by the above two regions of enhancement.

The formulas of the decomposed amplitudes and the probabilities derived in these two sections~\ref{sec:helio-perturbation-theory} and \ref{sec:DMP} may be utilized in analyses of the ongoing and the upcoming LBL experiments \cite{Abe:2019vii,Adamson:2020ypy,Acero:2019ksn,Abe:2018uyc,Abi:2020evt,Abe:2016ero}. In DUNE~\cite{Abi:2020evt}, the DMP decomposition might be more profitable because of its wide band beam, and it must be the choice to analyze the atmospheric neutrino observation~\cite{Abe:2017aap,Abe:2018uyc,Abi:2020evt,Abe:2016ero,TheIceCube-Gen2:2016cap,Adrian-Martinez:2016zzs}. In JUNO~\cite{An:2015jdp}, since it measures both the solar- and the atmospheric-scale oscillations, the DMP decomposition is the unique choice among the frameworks discussed in this paper. Though the matter effect in JUNO is small, $\sim$1\% level, it must be taken into account when the accuracy of measurement goes down to a percent level. We must emphasize, however, that to place the real significance to the above phenomenological prospects, we need to go through the analyses to prove the expectations mentioned in the last paragraph above. 

\section{Concluding remarks}
\label{sec:conclusion}

In this paper, we have addressed the question of how the amplitude decomposition can be defined in matter, the prescription of how to decompose the oscillation $S$ matrix into the ``atmospheric'' and ``solar'' amplitudes. In general, there are two qualitatively new features in neutrino oscillation in matter. Namely, the eigenvalues are modified by the matter effect, and the mixed mode of the $\Delta m^2_{31}$-driven and $\Delta m^2_{21}$-driven oscillations is generated even under a tiny matter potential. Therefore, generally, there is no well defined way in matter of decomposing the $S$ matrix into the atmospheric and solar waves in the same way as done in vacuum.

To know whether it is impossible or there is a way of circumventing the difficulty, we first tried an extension of our vacuum definition of amplitude decomposition into that in matter. We have found a successful case, the first-order AKS perturbation theory, which utilizes the hierarchy of the two $\Delta m^2$, $\epsilon \equiv \Delta m^2_{21} / \Delta m^2_{31} \ll 1$, and the weak matter effect. The region of validity may correspond to low energy of $\sim$ several 100 MeV and medium baselines of a few $\times 100$ km, which would be realized by T2K, T2HK, and ESS$\nu$SB settings. It offers probably the cleanest place for the amplitude decomposition in matter due to the unmodified vacuum frequencies of the two modes. 

Though finding the above specific example in which the vacuum definition works in matter is intriguing, it appears that the first-order AKS framework is the unique case, and generically a departure from the vacuum definition is necessary. It is because the energy eigenvalues and the effective mixing angles are modified in matter, behave dynamically, sometimes displaying a dramatic behavior. In this way the characterization, or what is implied by the ``atmospheric'' or the ``solar'' oscillations, can be obscured. 

To proceed toward treatment of amplitude decomposition in more generic kinematical phase space, we combined the two strategies:  
\begin{itemize}
\item
A formal definition of amplitude decomposition in matter, the Zaglauer-Schwarzer decomposition,

\item 
Analyzing the perturbative schemes in which the nature of the matter-modified atmospheric and solar oscillations are well understood.

\end{itemize}

One could hope that difficulties in understanding the physical properties of the decomposed two dynamical modes in the ZS definition are somehow cured at least partly in this way. We have analyzed so called the solar-resonance perturbation theory and the helio perturbation theory, which are discussed in sections~\ref{sec:solar-resonance-P} and \ref{sec:helio-perturbation-theory}, respectively. They are chosen due to their regions of validity, around the solar-scale and the atmospheric-scale enhanced oscillations, respectively. The necessity of implementing the general structure \'a la Zaglauer and Schwarzer became clear during the treatment of the solar-resonance perturbation theory. Integrating the lessons learned in these exercises, we were able to give the amplitude decomposition formulas in these perturbative frameworks.

In most part of this paper we have restricted ourselves into the $\nu_{\mu} \rightarrow \nu_{e}$ channel. It is because we concentrated on the conceptual issues, and our primary focus is on the question of what is the correct way of performing the amplitude decomposition, an indispensable tool in our approach. Of course, we must derive a complete set of the formulas for the decomposed probabilities of all the relevant oscillation channels toward the data analyses to extract and discuss the interference effects. This task will be carried out after we elevate the plausibility argument for the desired properties of the DMP framework given in section~\ref{sec:DMP-decomposition} to the solid results by explicit calculations. 

During the course of investigation in sections~\ref{sec:solar-resonance-P} and \ref{sec:helio-perturbation-theory}, we have found a new picture of the $\varphi$- and $\phi$-symmetries in the solar-resonance and the helio perturbation theories, respectively, as the ``protecting symmetries'' for the quantum mechanical interference. In spite of the existence of the symmetries, however, we have argued that the analysis procedure we propose with $\chi^2 (q)$ is tenable. 

The interference term in the probability reveals an interesting feature that its property, i.e., nature of the term, is oscillation-channel dependent. In the $\nu_{\mu} \rightarrow \nu_{e}$ channel the terms with CP phase $\delta$ are dominant. But, in the $\nu_{\mu} \rightarrow \nu_{\tau}$ (or $\nu_{\mu} \rightarrow \nu_{\mu}$) channel, the $\delta$-independent terms constitute the major component. The feature is true both in vacuum and in matter. Since experimental analyses with the LBL accelerator experiments may be more feasible with the $\nu_{\mu} \rightarrow \nu_{e}$ channel, measurement accuracy has to be sufficiently high if we want to show that the interference is not just due to CP phase effect but there is a $\delta$-independent contribution. In this context, the importance of the high-statistics LBL experiments, T2HK and DUNE, must be stressed for their greater capabilities for precision measurement. Yet, the analyses of T2K and NO$\nu$A data must be pursuit first to observe the interference term and to test the framework itself.

Finally, we have also reported our investigation of the amplitude decomposition in wider kinematical phase space using the DMP framework. 
We have established the DMP decomposition formulas by relying on the formulas given in the original reference. With incorporating the ZS structure, the decomposed amplitudes allows more physically appealing interpretation with the matter-dressed two mixing angles of $\theta_{13}$ and $\theta_{12}$. It is likely that the DMP framework interpolates the regions of validity of both of the solar- and atmospheric-resonance perturbation theories. If this is established the DMP framework can provide the appropriate method for the amplitude decomposition in matter, possibly in the whole kinematical region relevant for the atmospheric neutrino observation and the LBL experiments.

It is interesting to discuss physical picture outside the region of the matter-dressed atmospheric and solar variables in the context of amplitude decomposition. Identifying the nature of the two dynamical modes of oscillation would be easier at high energies, or high matter densities, because of dominance of one frequency. It may allow unified amplitude decomposition and interference analyses of low to super-high energy atmospheric neutrino observation by IceCube \cite{Aartsen:2020fwb} and the lower energy apparatus.

\acknowledgments

The author expresses deep gratitude to Pilar Coloma for useful discussions on amplitude decomposition beyond the perturbative regime and the cause of interference in the $\nu_\mu \rightarrow \nu_e$ appearance channel. 
He thanks Patrick Huber and Rebekah Pestes for exciting collaborations in the previous work \cite{Huber:2019frh}, the first publication on this topic, and interesting discussions in the early stage of this work. We have enjoyed great spiritual pressure from some of the pioneers who worked on this and the closely related subjects, in particular, Alexei Smirnov. Some of them and the various other people sent us their encouraging comments, in explicit or implicit manners. They include the ones from Samoil Bilenky, Sandhya Choubey, John Learned, Orlando Peres, Serguey Petcov, and Masashi Yokoyama. 

\appendix

\section{Matter perturbation theory of the three-flavor neutrino oscillation}
\label{sec:matter-P-theory}

To formulate the matter perturbation theory we transform from the flavor basis to the vacuum mass eigenstate basis, the check basis 
\begin{eqnarray} 
\check{\nu}_{\alpha} = U^{\dagger}_{\alpha \beta} \nu_{\beta}
= \left( U_{23} U_{13} U_{12} \right)^{\dagger}_{\alpha \beta} \nu_{\beta}, 
\label{check-basis}
\end{eqnarray}
with the Hamiltonian 
\begin{eqnarray} 
\check{H} 
&=& 
\left( U_{23} U_{13} U_{12} \right)^{\dagger} H U_{23} U_{13} U_{12}
%
%\nonumber \\&=&
= 
\left[
\begin{array}{ccc}
0 & 0 & 0 \\
0 & \Delta_{21} & 0 \\
0 & 0 & \Delta_{31} \\
\end{array}
\right] 
+ U_{12}^{\dagger} U_{13}^{\dagger} 
\left[
\begin{array}{ccc}
\Delta_{a} & 0 & 0 \\
0 & 0 & 0 \\
0 & 0 & 0 \\
\end{array}
\right] 
U_{13} U_{12} 
\nonumber \\
&=&
\left[
\begin{array}{ccc}
0 & 0 & 0 \\
0 & \Delta_{21} & 0 \\
0 & 0 & \Delta_{31} \\
\end{array}
\right] 
+ 
\left[
\begin{array}{ccc}
c^2_{12} c^2_{13} \Delta_{a} & 
c_{12} s_{12} c^2_{13} \Delta_{a} & 
c_{12}c_{13} s_{13} e^{ -i \delta } \Delta_{a} \\
c_{12} s_{12} c^2_{13} \Delta_{a} & 
s^2_{12} c^2_{13} \Delta_{a} & 
s_{12} c_{13} s_{13} e^{ -i \delta } \Delta_{a} \\
c_{12} c_{13} s_{13} e^{ i \delta } \Delta_{a} & 
s_{12} c_{13} s_{13} e^{ i \delta } \Delta_{a} & 
s^2_{13} \Delta_{a} \\
\end{array}
\right], 
\label{check-H-def}
\end{eqnarray}
where $\Delta_{a} \equiv \frac{ a }{ 2E }$ as defined in eq.~\eqref{Delta-a-def}. We denote the first and second terms in \eqref{check-H-def} the unperturbed and perturbed Hamiltonians in the check basis, respectively. 

A conventional perturbative treatment entails the expressions of the $\check{S}$ matrix to first order as 
\begin{eqnarray} 
\check{S} (x) =  
e^{-i \check{H}_{0} x} %% \Omega(x). 
\left[
1 + (-i) \int^{x}_{0} dx' H_{1} (x') 
\right], 
\label{check-Smatrix-def} 
\end{eqnarray}
where 
\begin{eqnarray} 
&& 
H_{1} \equiv e^{i \check{H}_{0} x} \check{H}_{1} e^{-i \check{H}_{0} x}. 
\label{H1-def}
\end{eqnarray}
The explicit expressions of zeroth and first order $\check{S}$ matrix elements in the check basis, $\check{S} (x) = \check{S}^{(0)} (x) + \check{S}^{(1)} (x)$, can be written as 
\begin{eqnarray} 
&& 
\check{S}^{(0)} (x) 
= 
\left[
\begin{array}{ccc}
e^{ - i h_{1} x } & 0 & 0 \\
0 & e^{ - i h_{2} x } & 0 \\
0 & 0 & e^{ - i h_{3} x } \\
\end{array}
\right], 
\nonumber \\
&& 
\check{S}^{(1)} (x) 
\nonumber \\
&& \hspace{-24mm} 
=
\left[
\begin{array}{ccc}
c^2_{12} c^2_{13} ( -i \Delta_{a} x) e^{ - i h_{1} x } & 
c_{12} s_{12} c^2_{13} \frac{ \Delta_{a} }{ h_{2} - h_{1} } 
\left\{ e^{ - i h_{2} x } - e^{ - i h_{1} x }  \right\} & 
c_{12} c_{13} s_{13} e^{ -i \delta } 
\frac{ \Delta_{a} }{ h_{3} - h_{1} } 
\left\{ e^{ - i h_{3} x } - e^{ - i h_{1} x } \right\} \\
c_{12} s_{12} c^2_{13} \frac{ \Delta_{a} }{ h_{2} - h_{1} } 
\left\{ e^{ - i h_{2} x } - e^{ - i h_{1} x }  \right\} & 
s^2_{12} c^2_{13} ( -i \Delta_{a} x) e^{ - i h_{2} x } & 
s_{12} c_{13} s_{13} e^{ -i \delta } 
\frac{ \Delta_{a} }{ h_{3} - h_{2} } 
\left\{ e^{ - i h_{3} x } - e^{ - i h_{2} x } \right\} \\
c_{12} c_{13} s_{13} e^{ i \delta } 
\frac{ \Delta_{a} }{ h_{3} - h_{1} } 
\left\{ e^{ - i h_{3} x } - e^{ - i h_{1} x } \right\} & 
s_{12} c_{13} s_{13} e^{ i \delta } 
\frac{ \Delta_{a} }{ h_{3} - h_{2} } 
\left\{ e^{ - i h_{3} x } - e^{ - i h_{2} x } \right\} & 
s^2_{13} ( -i \Delta_{a} x) e^{ - i h_{3} x } \\
\end{array}
\right], 
\nonumber \\
\label{check-Smatrix} 
\end{eqnarray}
where $h_{i}$ ($i=1,2,3$) denote the eigenvalues of $\check{H}_{0}$ and in our case they are the as in vacuum: $h_{1} = 0$, $h_{2} = \Delta_{21}$, and $h_{3} = \Delta_{31}$.

Then, the flavor basis $S$ matrix can be obtained as 
\begin{eqnarray}
&& S = U \check{S} U^{\dagger}
= U_{23} U_{13} U_{12} \check{S} \left( U_{23} U_{13} U_{12} \right)^{\dagger}. 
\label{flavor-Smatrix}
\end{eqnarray}
Using the flavor basis $S$ matrix element the oscillation probability $P(\nu_{\beta} \rightarrow \nu_{\alpha})$ is given by 
\begin{eqnarray}
&& 
P(\nu_{\beta} \rightarrow \nu_{\alpha}) 
= \vert S_{\alpha \beta} \vert^2.
\label{oscillation-probability-def}
\end{eqnarray}

\section{Formulation of the AKS perturbation theory}
\label{sec:formulation-AKS}

In the AKS perturbation theory, in which we use the check basis as in appendix~\ref{sec:matter-P-theory}, we use a different decomposition of the vacuum mass eigenstate basis Hamiltonian into the unperturbed and perturbed parts as 
\begin{eqnarray}
\check{H} &=& \check{H}_{0} + \check{H}_{1}, 
\hspace{14mm} 
\check{H}_{0} = 
\left[
\begin{array}{ccc}
0 & 0 & 0 \\
0 & 0 & 0 \\
0 & 0 & \Delta_{31} \\
\end{array}
\right], 
\nonumber \\ 
\check{H}_{1} &=& 
\left[
\begin{array}{ccc}
0 & 0 & 0 \\
0 & \Delta_{21} & 0 \\
0 & 0 & 0 \\
\end{array}
\right] 
+ 
\left[
\begin{array}{ccc}
c^2_{12} c^2_{13} \Delta_{a} & 
c_{12} s_{12} c^2_{13} \Delta_{a} & 
c_{12}c_{13} s_{13} e^{ -i \delta } \Delta_{a} \\
c_{12} s_{12} c^2_{13} \Delta_{a} & 
s^2_{12} c^2_{13} \Delta_{a} & 
s_{12} c_{13} s_{13} e^{ -i \delta } \Delta_{a} \\
c_{12} c_{13} s_{13} e^{ i \delta } \Delta_{a} & 
s_{12} c_{13} s_{13} e^{ i \delta } \Delta_{a} & 
s^2_{13} \Delta_{a} \\
\end{array}
\right].
\label{check-H-AKS}
\end{eqnarray}
That is, not only the matter potential but also the $\Delta_{21}$ terms are assumed to be small, anticipating use of the formulas in regions of atmospheric-scale enhanced oscillation, $\Delta m^2_{31} L / 4E \sim \mathcal{O} (1)$, at short or medium baseline $L \simeq$ a few $\times$100 km. It corresponds, for example, to the T2K \cite{Abe:2019vii}, T2HK \cite{Abe:2018uyc}, and ESS$\nu$SB \cite{Baussan:2013zcy} experiments. 
Since the zeroth order Hamiltonian $\check{H}_{0}$ is diagonal one can do perturbative calculation in this basis.\footnote{
%%%%%%%%%%%%% footnote %%%%%%%%%%%%%
In ref.~\cite{Arafune:1997hd} the authors takes a different way by saying that they do perturbative calculation in the flavor basis, but in net what they do is the same as we explain here. 
}
Following the description of how perturbative expansion is organized in the previous section, we just present the results of the $\check{S}$ matrix elements. The zeroth order $\check{S}^{(0)}$ matrix is the same as in \eqref{check-Smatrix}, but now $h_{1} = 0, h_{2} = 0, h_{3} = \Delta_{31}$. The first order $\check{S}^{(1)}$ matrix is given by 
\begin{eqnarray} 
&& 
\check{S}^{(1)} = 
\left[
\begin{array}{ccc}
0 & 0 & 0 \\
0 & (-i \Delta_{21} x) & 0 \\
0 & 0 & 0 
\end{array}
\right] 
\nonumber \\
&+&
\left[
\begin{array}{ccc}
c^2_{12} c^2_{13} (-i \Delta_{a} x) & 
c_{12} s_{12} c^2_{13} (-i \Delta_{a} x) & 
- c_{12}c_{13} s_{13} e^{ -i \delta } \Delta_{a} 
\frac{ 1 - e^{ -i \Delta_{31} x } }{ \Delta_{31} } 
\\
c_{12} s_{12} c^2_{13} (-i \Delta_{a} x) & 
s^2_{12} c^2_{13} (-i \Delta_{a} x) & 
- s_{12} c_{13} s_{13} e^{ -i \delta } \Delta_{a} 
\frac{ 1 - e^{ -i \Delta_{31} x } }{ \Delta_{31} }  \\
- c_{12} c_{13} s_{13} e^{ i \delta } \Delta_{a} 
\frac{ 1 - e^{ -i \Delta_{31} x } }{ \Delta_{31} } & 
- s_{12} c_{13} s_{13} e^{ i \delta } \Delta_{a} 
\frac{ 1 - e^{ -i \Delta_{31} x } }{ \Delta_{31} } & 
s^2_{13} (-i \Delta_{a} x) e^{ -i \Delta_{31} x } \\
\end{array}
\right].
\nonumber \\
\label{check-Smatrix-1st}
\end{eqnarray}

Then, the flavor basis $S$ matrix can readily be calculated by using the formula in \eqref{flavor-Smatrix}, $S = U \check{S} U^{\dagger}$. 
The zeroth order $S$ matrix reads has the familiar vacuum form 
\begin{eqnarray} 
&& 
S^{(0)} (x) 
= 
U \left[
\begin{array}{ccc}
1 & 0 & 0 \\
0 & 1 & 0 \\
0 & 0 & e^{ -i x \Delta_{31} }
\end{array}
\right] U^{\dagger} 
\nonumber \\
&=&
\left[
\begin{array}{ccc}
1 + \vert U_{13} \vert^2 \left( e^{ -i x \Delta_{31} } - 1 \right) & 
U_{13} U_{23}^* \left( e^{ -i x \Delta_{31} } - 1 \right) & 
U_{13} U_{33} \left( e^{ -i x \Delta_{31} } - 1 \right) \\
U_{23} U_{13}^* \left( e^{ -i x \Delta_{31} } - 1 \right) & 
1 + \vert U_{23} \vert^2 \left( e^{ -i x \Delta_{31} } - 1 \right) & 
U_{23} U_{33} \left( e^{ -i x \Delta_{31} } - 1 \right) \\
U_{33} U_{13}^* \left( e^{ -i x \Delta_{31} } - 1 \right) & 
U_{33} U_{23}^* \left( e^{ -i x \Delta_{31} } - 1 \right) & 
1 + U^2_{33} \left( e^{ -i x \Delta_{31} } - 1 \right) \\
\end{array}
\right]. 
\label{Smatrix-zeroth-def}
\end{eqnarray}
We denote the two parts of the first order $S^{(1)}$ matrix as $S^{(1)} = S^{(1)}_{ \text{helio} } + S^{(1)}_{ \text{matter} }$, which come from the two different components of $\check{H}_{1}$ in \eqref{check-H-AKS}. $S^{(1)}_{ \text{helio} }$ reads  
\begin{eqnarray}
&& 
S^{(1)}_{ \text{helio} } 
%\nonumber \\&=& 
= ( - i \Delta_{21} x) 
\left[
\begin{array}{ccc}
\vert U_{12} \vert^2 & 
U_{12} U_{22}^* & 
U_{12} U_{32}^* \\
U_{22} U_{12}^* & U^2_{22} & U_{22} U_{32}^* \\
U_{32} U_{12}^* & U_{32} U_{22} & \vert U_{32} \vert^2 \\
\end{array}
\right], 
\end{eqnarray}
while the expression of the elements of $S^{(1)}_{ \text{matter} }$ is a little more cumbersome. But, they can be calculated in a straightforward manner by using the formula 
\begin{eqnarray} 
S^{(1)}_{ \text{matter} } = U \check{S}^{(1)} U^{\dagger},
\label{S-matter-1st} 
\end{eqnarray} 
whose $e - \mu$ element is given in eq.~\eqref{S-1st-matter-AKS-1}. 

\section{Formulation of the helio perturbation theory}
\label{sec:formulation-helio-perturbation}

Here, we review the formulation of the helio perturbation theory to recollect the necessary formulas for discussion of the amplitude decomposition in section~\ref{sec:helio-perturbation-theory}.

\subsection{Tilde basis and diagonalization of the zeroth-order Hamiltonian}
\label{sec:tilde-basis}

To formulate the helio perturbation theory with the unique expansion parameter $\epsilon$ as defined in \eqref{epsilon-def}, we use the tilde basis $\tilde{\nu}_{\alpha} = (U_{23}^{\dagger})_{\alpha \beta} \nu_{\beta}$ and $\tilde{H} = U_{23}^{\dagger} H U_{23}$. Starting from the one in the vacuum mass eigenstate basis \eqref{check-H-def}, the Hamiltonian in the tilde basis is obtained as 
\begin{eqnarray}
\tilde{H} 
&=& 
U_{13} U_{12} \check{H} U_{12}^{\dagger} U_{13}^{\dagger}.
\label{tilde-H-def}
\end{eqnarray}
The tilde basis Hamiltonian is decomposed into unperturbed and perturbed part as 
\begin{eqnarray}
\tilde{H} &=& \tilde{H}_{0} + \tilde{H}_{1}, 
\nonumber \\ 
\tilde{H}_{0} &=&
\Delta_{31}
\left[
\begin{array}{ccc}
s^2_{13} + r_{a} & 0 & c_{13} s_{13} e^{ - i \delta} \\
0 & 0 & 0 \\
c_{13} s_{13} e^{ i \delta} & 0 & c^2_{13} \\
\end{array}
\right], 
\nonumber \\ 
\tilde{H}_{1} &=& 
\Delta_{21}
\left[
\begin{array}{ccc}
c^2_{13} s^2_{12} & 
c_{13} c_{12} s_{12} & 
- c_{13} s_{13} s^2_{12} e^{- i \delta} \\
c_{13} c_{12} s_{12} & 
c^2_{12} & 
- s_{13} c_{12} s_{12} e^{ - i \delta}  \\
- c_{13} s_{13} s^2_{12} e^{ i \delta} & 
- s_{13} c_{12} s_{12} e^{ i \delta} & 
s^2_{13} s^2_{12}  \\
\end{array}
\right] \equiv \Delta_{21} F.
\label{tilde-H-F-def}
\end{eqnarray}
where we have defined the $F$ matrix and introduced $r_{a}$, the matter to vacuum ratio, 
\begin{eqnarray} 
r_{a} \equiv \frac{a}{ \Delta m^2_{31} } 
= \frac{ \Delta_{a} }{ \Delta_{31} }. 
\label{matter-vacuum-ratio}
\end{eqnarray}
As in \eqref{tilde-H-F-def}, the $F$ matrix is simply $\Delta_{21}$ scaled $\tilde{H}_{1}$. The general expression of $\tilde{H}_{1}$ with the $F_{ij}$ $(i,j=1,2,3)$ elements will help us to understand the general features of the theory. See appendix~\ref{sec:1st-order}. 

The zeroth-order tilde-basis Hamiltonian $\tilde{H}_{0}$ can be easily diagonalized by unitary transformation $U_{\phi}$ parametrized as 
\begin{eqnarray} 
U_{\phi} =
\left[
\begin{array}{ccc}
c_{\phi} & 0 & s_{\phi} e^{ - i \delta} \\
0 & 1 & 0 \\
- s_{\phi} e^{ i \delta} & 0 & c_{\phi} 
\end{array}
\right] 
\label{eq:U-phi}
\end{eqnarray}
such that 
\begin{eqnarray} 
\hat{H}_{0} &=& U^{\dagger}_{\phi} \tilde{H}_{0} U_{\phi} 
= 
\left[
\begin{array}{ccc}
h_{1} & 0 & 0 \\
0 & h_{2} & 0 \\
0 & 0 & h_{3} 
\end{array}
\right].
\label{hat-hamiltonian-0th}
\end{eqnarray}
The diagonalization determines $\phi$, the angle $\theta_{13}$ in matter, as 
\begin{eqnarray} 
&&
\cos 2\phi 
= \frac{ \cos 2\theta_{13} - r_{a} }{ \sqrt{ ( \cos 2\theta_{13} - r_{a} )^2 + \sin^2 2\theta_{13} } }, 
\nonumber \\
&&
\sin 2\phi 
= \frac{ \sin 2\theta_{13} }{ \sqrt{ ( \cos 2\theta_{13} - r_{a} )^2 + \sin^2 2\theta_{13} } }. 
\label{cos-sin-2phi} 
\end{eqnarray}
We hereafter denote the basis \eqref{hat-hamiltonian-0th}, the $\tilde{H}_{0}$ diagonalized basis, as the hat basis with notation $\hat{H}_{0}$. The eigenvalues $h_{i}$ are given by
\begin{eqnarray} 
h_{1} &=& \sin^2 ( \phi - \theta_{13} ) \Delta_{31} 
+ c^2_{\phi} \Delta_{a} 
= \frac{ \Delta_{31} }{2} \left[ 1 + r_{a} \mp \sqrt{ 1 + r_{a}^2 - 2 r_{a} \cos 2\theta_{13} } \right], 
\nonumber\\
h_{2} &=& 0, 
\nonumber\\
h_{3} &=& \cos^2 ( \phi - \theta_{13} ) \Delta_{31} 
+ s^2_{\phi} \Delta_{a} 
= \frac{ \Delta_{31} }{2} \left[ 1 + r_{a} \pm \sqrt{ 1 + r_{a}^2 - 2 r_{a} \cos 2\theta_{13} } \right], 
\label{eigenvalues}
\end{eqnarray}
where the upper and lower signs correspond to the normal and inverted mass orderings ~\cite{Minakata:2015gra}.\footnote{
%%%%%%%%%%%% footnote %%%%%%%%%%%%%%
Notice that our state labels are defined by the zeroth-order Hamiltonian in eq.~\eqref{tilde-H-F-def}. In the case of normal mass ordering, the ordering of the eigenvalues are such that $h_{2} < h_{1} \ll h_{3}$ ($h_{1} \ll h_{2} < h_{3}$) in the $a \rightarrow + \infty$ ($a \rightarrow - \infty$) limit, and $h_{2} = h_{1} <  h_{3}$ in vacuum. Therefore, the atmospheric resonance exists between the eigenstates 3 and 1, and our state labeling is different from the conventional one in which the resonance is between eigenstates 3 and 2. It stems from the fact that the solar level crossing is not treated properly, the inherent problem in the formulation of the helio perturbation theory so far presented~\cite{Arafune:1996bt,Cervera:2000kp,Freund:2001pn,Akhmedov:2004ny,Minakata:2015gra}, as discussed in ref.~\cite{Minakata:2015gra}.
}
The two expressions of $h_{1}$ and $h_{3}$ are both valid.

\subsection{$\hat{S}$ matrix in the hat basis vs. $S$ matrix in the flavor basis}
\label{sec:basis-relations}

The relationship between the various basis:  
\begin{eqnarray}
&& \tilde{H} = U_{23}^{\dagger} H U_{23}, 
\hspace{10mm}
\hat{H} = U_{\phi}^{\dagger} \tilde{H} U_{\phi}, 
\hspace{10mm}
\tilde{H} = U_{\phi} \hat{H} U_{\phi}^{\dagger}, 
\nonumber \\
&& 
H = U_{23} \tilde{H} U_{23}^{\dagger} 
= U_{23} U_{\phi} \hat{H} U_{\phi}^{\dagger} 
U_{23}^{\dagger}. 
\end{eqnarray}
The last relation applies to the $S$ matrix as well 
\begin{eqnarray}
&& 
S 
= U_{23} \tilde{S} U_{23}^{\dagger} 
= U_{23} U_{\phi} \hat{S} U_{\phi}^{\dagger} U_{23}^{\dagger}. 
\label{hatS-S-relation}
\end{eqnarray}

\subsection{The zeroth order $\hat{S}$ and $S$ matrices }
\label{sec:S-matrix-0th}

Let us calculate first the flavor basis $S$ matrix in the zeroth order. The hat basis $S$ matrix in the zeroth order is given by 
\begin{eqnarray} 
\hat{S}^{(0)} &=&  
\left[
\begin{array}{ccc}
e^{ - i h_{1} x } & 0 & 0 \\
0 & e^{ - i h_{2} x } & 0 \\
0 & 0 & e^{ - i h_{3} x } 
\end{array}
\right]. 
\label{hat-hamiltonian1}
\end{eqnarray}
Then by performing the $U_{\phi}$ and $U_{23} U_{\phi}$ rotations we obtain 
\begin{eqnarray} 
&& 
\tilde{S}^{(0)} = U_{\phi} \hat{S} U_{\phi}^{\dagger} 
\nonumber \\
&=& 
\left[
\begin{array}{ccc}
s^2_{\phi} e^{ - i h_{3} x } + c^2_{\phi} e^{ - i h_{1} x } & 
0 & 
c_{\phi} s_{\phi} e^{ - i \delta} \left( e^{ - i h_{3} x } - e^{ - i h_{1} x } \right) \\
0 & e^{ - i h_{2} x } & 0 \\
c_{\phi} s_{\phi} e^{ i \delta} \left( e^{ - i h_{3} x } - e^{ - i h_{1} x } \right) & 
0 & 
c^2_{\phi} e^{ - i h_{3} x } + s^2_{\phi} e^{ - i h_{1} x }
\end{array}
\right], 
\end{eqnarray}
\begin{eqnarray}
&& 
S^{(0)} = U_{23} \tilde{S}^{(0)} U_{23}^{\dagger} 
\nonumber \\
&&
\hspace{-20mm} 
=
%\nonumber \\&=& 
\left[
\begin{array}{ccc}
s^2_{\phi} e^{ - i h_{3} x } + c^2_{\phi} e^{ - i h_{1} x } & 
s_{23} c_{\phi} s_{\phi} e^{ - i \delta} \left( e^{ - i h_{3} x } - e^{ - i h_{1} x } \right) & 
c_{23} c_{\phi} s_{\phi} e^{ - i \delta} \left( e^{ - i h_{3} x } - e^{ - i h_{1} x } \right) \\
s_{23} c_{\phi} s_{\phi} e^{ i \delta} \left( e^{ - i h_{3} x } - e^{ - i h_{1} x } \right) & 
s^2_{23} \left( c^2_{\phi} e^{ - i h_{3} x } + s^2_{\phi} e^{ - i h_{1} x } \right) 
+ c^2_{23} e^{ - i h_{2} x } & 
c_{23} s_{23} \left( c^2_{\phi} e^{ - i h_{3} x } + s^2_{\phi} e^{ - i h_{1} x } - e^{ - i h_{2} x } \right) \\
c_{23} c_{\phi} s_{\phi} e^{ i \delta} \left( e^{ - i h_{3} x } - e^{ - i h_{1} x } \right) & 
c_{23} s_{23} \left( c^2_{\phi} e^{ - i h_{3} x } + s^2_{\phi} e^{ - i h_{1} x } - e^{ - i h_{2} x } \right) & 
c^2_{23} \left( c^2_{\phi} e^{ - i h_{3} x } + s^2_{\phi} e^{ - i h_{1} x } \right) 
+ s^2_{23} e^{ - i h_{2} x } 
\end{array}
\right].
\nonumber \\
\label{flavor-basis-Smatrix-0th}
\end{eqnarray}

\subsection{The first order correction} 
\label{sec:1st-order} 

The first order correction can be calculated  by using the formulas \eqref{check-Smatrix-def} and \eqref{H1-def}, but in the hat basis. Let us calculate $H_{1}$ first: 
\begin{eqnarray} 
&& 
H_{1} 
= e^{i \hat{H}_{0} x} \hat{H}_{1} e^{-i \hat{H}_{0} x} 
= e^{i \hat{H}_{0} x} U_{\phi}^{\dagger} \tilde{H}_{1} U_{\phi} e^{-i \hat{H}_{0} x} 
\nonumber \\
&=& 
U_{\phi}^{\dagger} 
\left( U_{\phi} e^{i \hat{H}_{0} x} U_{\phi}^{\dagger} \right) 
\tilde{H}_{1} 
\left( U_{\phi} e^{-i \hat{H}_{0} x} U_{\phi}^{\dagger} \right) 
U_{\phi} 
= \Delta_{21} U_{\phi}^{\dagger} \Phi U_{\phi}, 
\label{H1-result}
\end{eqnarray}
where the factors inside parentheses can be obtained as 
\begin{eqnarray}  
U_{\phi} e^{ \pm i \hat{H}_{0} x} U_{\phi}^{\dagger} 
&=&
\left[
\begin{array}{ccc}
s^2_{\phi} e^{ \pm i h_{3} x } + c^2_{\phi} e^{ \pm i h_{1} x } & 
0 & 
c_{\phi} s_{\phi} e^{ - i \delta} \left( e^{ \pm i h_{3} x } - e^{ \pm i h_{1} x } \right) \\
0 & e^{ \pm i h_{2} x } & 0 \\
c_{\phi} s_{\phi} e^{ i \delta} \left( e^{ \pm i h_{3} x } - e^{ \pm i h_{1} x } \right) & 
0 & 
c^2_{\phi} e^{ \pm i h_{3} x } + s^2_{\phi} e^{ \pm i h_{1} x }
\end{array}
\right] 
\end{eqnarray}
where it should be noticed that $U_{\phi} e^{ - i \hat{H}_{0} x} U_{\phi}^{\dagger} = \tilde{S}^{(0)}$. In \eqref{H1-result}, the $\Phi$ matrix is defined as 
\begin{eqnarray} 
&& 
\Phi 
\equiv 
\left( U_{\phi} e^{i \hat{H}_{0} x} U_{\phi}^{\dagger} \right) 
F 
\left( U_{\phi} e^{-i \hat{H}_{0} x} U_{\phi}^{\dagger} \right), 
\label{Phi-def}
\end{eqnarray}
where the $F$ matrix is defined in \eqref{tilde-H-F-def}. The computed results of the $\Phi$ matrix elements are given in appendix~\ref{sec:F-Phi-summary}. 
Then, the first order $\hat{S}$ matrix, and $\tilde{S}$ matrix are given, respectively as 
\begin{eqnarray} 
&& \hat{S}^{(1)} =  
\Delta_{21} U_{\phi}^{\dagger} 
\left( U_{\phi} e^{-i \hat{H}_{0} x} U_{\phi}^{\dagger} \right) 
\left[ (-i) \int^{x}_{0} dx' \Phi (x') \right]
U_{\phi},  
\nonumber \\
&& 
\tilde{S}^{(1)} = 
U_{\phi} \hat{S}^{(1)} U_{\phi}^{\dagger} 
%\nonumber \\&=& 
= \Delta_{21} 
\left( U_{\phi} e^{-i \hat{H}_{0} x} U_{\phi}^{\dagger} \right) 
\left[ (-i) \int^{x}_{0} dx' \Phi (x') \right].  
\label{hat-Smatrix}
\end{eqnarray}

Now, the knowledgeable readers might have noticed that our computation of the first order corrections is exactly parallel to that of the ``helio-UV perturbation theory'' formulated and discussed in ref.~\cite{Martinez-Soler:2018lcy}. This is true despite that the physical meaning of the correction terms is very different, the ``helio correction'' in our case and the unitarity violating effect in ref.~\cite{Martinez-Soler:2018lcy}. The correspondence is that our $\Delta_{21}$ is $\Delta_{b}$ (neutral current version of $\Delta_{a}$), and our $F$ matrix is the $H$ matrix in ref.~\cite{Martinez-Soler:2018lcy}. 
Minor differences are in the choice of convention of the flavor mixing matrix, the one in PDG convention \eqref{MNSmatrix-PDG} in the present paper, in contrast to the ATM convention in ref.~\cite{Martinez-Soler:2018lcy}. This correspondence can be used as a consistency check of the calculation. 

\subsection{Flavor basis $S$ matrix and the oscillation probability}  
\label{sec:S-matrix-probability} 

The flavor basis $S$ matrix is given by $S = U_{23} \tilde{S} U_{23}^{\dagger}$ as in \eqref{hatS-S-relation}. The relations of $\tilde{S}$ and $S$ matrix elements are explicitly written in eq.~\eqref{tildeS-S-relation} in appendix~\ref{sec:tilde-S(1)-summary}. Then, the oscillation probability $P(\nu_{\beta} \rightarrow \nu_{\alpha})$ is given by eq.~\eqref{oscillation-probability-def}. 

\section{$F$ and $\Phi$ matrix elements summary}
\label{sec:F-Phi-summary}

Here is the summary of the $F$ matrix elements defined in \eqref{tilde-H-F-def}, and the computed result of $\Phi$ matrix elements defined in \eqref{Phi-def} as a function of the $F$ matrix elements. The PDG convention of the flavor mixing matrix is used.
\begin{eqnarray} 
&& F_{11} = c^2_{13} s^2_{12}, 
\nonumber \\
&& F_{12} = c_{13} c_{12} s_{12} = F_{21}, 
\nonumber \\
&& F_{13} = - c_{13} s_{13} s^2_{12} e^{- i \delta} = \left( F_{31} \right)^*,
\nonumber \\
&& F_{22} = c^2_{12}, 
\nonumber \\
&& F_{23} = - s_{13} c_{12} s_{12} e^{ - i \delta} = \left( F_{32} \right)^*,
\nonumber \\
&& F_{33} = s^2_{13} s^2_{12}.
\label{Fij-elements}
\end{eqnarray}
\begin{eqnarray} 
\Phi_{11} &=& 
F_{11} + 
c_{\phi} s_{\phi} \left[ \sin 2\phi 
( F_{33} - F_{11} )
- \cos 2 \phi 
\left( e^{ i \delta} F_{13} + e^{ - i \delta} F_{31} \right) \right]
\nonumber \\
&-& 
e^{ - i ( h_{3} -  h_{1} ) x } 
c_{\phi} s_{\phi} 
\left[ c_{\phi} s_{\phi} ( F_{33} - F_{11} )
- \left( c^2_{\phi} e^{ i \delta} F_{13} - s^2_{\phi} e^{ - i \delta} F_{31} \right) 
\right] 
\nonumber \\
&-& 
e^{ i ( h_{3} -  h_{1} ) x } 
c_{\phi} s_{\phi} 
\left[ c_{\phi} s_{\phi} ( F_{33} - F_{11} ) 
+ \left( s^2_{\phi} e^{ i \delta} F_{13} - c^2_{\phi} e^{ - i \delta} F_{31} \right) 
\right], 
\nonumber \\
\Phi_{12} &=& 
\left( s^2_{\phi} F_{12} + c_{\phi} s_{\phi} e^{ - i \delta} F_{32} \right) 
e^{ i ( h_{3} -  h_{2} ) x } 
+ \left( c^2_{\phi} F_{12} - c_{\phi} s_{\phi} e^{ - i \delta} F_{32} \right) 
e^{ - i ( h_{2} -  h_{1} ) x }, 
\nonumber \\
\Phi_{13} &=& 
e^{ - i \delta} \biggl\{ 
c_{\phi} s_{\phi} \left[ 
\cos 2 \phi ( F_{33} - F_{11} ) 
+ \sin 2\phi 
\left( e^{ i \delta} F_{13} + e^{ - i \delta} F_{31} \right) 
\right] 
\nonumber \\
&+& 
e^{ - i ( h_{3} -  h_{1} ) x } 
c^2_{\phi} \left[ 
- c_{\phi} s_{\phi} ( F_{33} - F_{11} ) 
+ c^2_{\phi} e^{ i \delta} F_{13} - s^2_{\phi} e^{ - i \delta} F_{31}
\right] 
\nonumber \\
&+& 
e^{ i ( h_{3} -  h_{1} ) x } 
s^2_{\phi} \left[ 
c_{\phi} s_{\phi} ( F_{33} - F_{11} ) 
+ s^2_{\phi} e^{ i \delta} F_{13} - c^2_{\phi} e^{ - i \delta} F_{31}
\right]
\biggr\}.
\end{eqnarray}
\begin{eqnarray} 
\Phi_{21} 
&=& 
e^{ - i ( h_{3} - h_{2} ) x } 
\left( s^2_{\phi} F_{21} + c_{\phi} s_{\phi} e^{ i \delta} F_{23} \right) 
+ e^{ i ( h_{2} - h_{1} ) x } 
\left( c^2_{\phi} F_{21} - c_{\phi} s_{\phi} e^{ i \delta} F_{23} \right), 
\nonumber \\
\Phi_{22} &=& F_{22}, 
\nonumber \\
\Phi_{23} 
&=& 
 e^{ - i ( h_{3} - h_{2} ) x } 
 \left( c_{\phi} s_{\phi} e^{ - i \delta} F_{21} + c^2_{\phi} F_{23} \right) 
 - e^{ i ( h_{2} - h_{1} ) x } 
 \left( c_{\phi} s_{\phi} e^{ - i \delta} F_{21} - s^2_{\phi} F_{23} \right).
\end{eqnarray}
\begin{eqnarray} 
\Phi_{31} 
&=& 
e^{ i \delta} \biggl\{ 
c_{\phi} s_{\phi} 
\left[ \cos 2\phi ( F_{33} - F_{11} ) 
+ \sin 2\phi    %% 2 c_{\phi} s_{\phi} 
\left( e^{ i \delta} F_{13} + e^{ - i \delta} F_{31} \right) \right] 
\nonumber \\
&+& 
e^{ - i ( h_{3} -  h_{1} ) x } s^2_{\phi} 
\left[ 
c_{\phi} s_{\phi} ( F_{33} - F_{11} ) 
- \left( c^2_{\phi} e^{ i \delta} F_{13} - s^2_{\phi} e^{ - i \delta} F_{31} \right) 
\right]
\nonumber \\
&-& 
e^{ i ( h_{3} -  h_{1} ) x } c^2_{\phi} 
\left[
c_{\phi} s_{\phi} ( F_{33} - F_{11} ) 
+ \left( s^2_{\phi} e^{ i \delta} F_{13} - c^2_{\phi} e^{ - i \delta} F_{31} \right) 
\right]
\biggr\}, 
\nonumber \\
\Phi_{32} &=& 
e^{ i ( h_{3} - h_{2} ) x } 
\left( c_{\phi} s_{\phi} e^{ i \delta} F_{12} + c^2_{\phi} F_{32} \right) 
- e^{ - i ( h_{2} - h_{1} ) x } 
\left( c_{\phi} s_{\phi} e^{ i \delta} F_{12} - s^2_{\phi} F_{32} \right), 
\nonumber \\
\Phi_{33} 
&=& 
F_{33} - 
c_{\phi} s_{\phi} 
\left[ \sin 2\phi ( F_{33} - F_{11} ) 
- \cos 2\phi 
\left( e^{ i \delta} F_{13} + e^{ - i \delta} F_{31} \right) 
\right]
\nonumber \\
&+& 
e^{ - i ( h_{3} -  h_{1} ) x } 
c_{\phi} s_{\phi} \left[
c_{\phi} s_{\phi} ( F_{33} - F_{11} ) 
- \left( c^2_{\phi} e^{ i \delta} F_{13} - s^2_{\phi} e^{ - i \delta} F_{31} \right) 
\right] 
\nonumber \\
&+& 
e^{ i ( h_{3} -  h_{1} ) x } 
c_{\phi} s_{\phi} \left[
c_{\phi} s_{\phi} ( F_{33} - F_{11} ) 
+ \left( s^2_{\phi} e^{ i \delta} F_{13} - c^2_{\phi} e^{ - i \delta} F_{31} \right) 
\right]. 
\end{eqnarray}

\section{$\tilde{S}^{(1)}$ matrix elements summary and $\tilde{S}$-$S$ matrix relation}
\label{sec:tilde-S(1)-summary} 

We present computed results of the first order $\tilde{S}^{(1)}$ matrix elements.
%
%%%%%%%%%%%% S_11 %%%%%%%%%%%%%%
\begin{eqnarray}
&& 
\tilde{S}^{(1)}_{11} 
\nonumber \\ 
&=& 
F_{11} 
\left( s^2_{\phi} e^{ - i h_{3} x } + c^2_{\phi} e^{ - i h_{1} x } \right) 
( - i \Delta_{21} x ) 
\nonumber \\ 
&+& 
c_{\phi} s_{\phi} 
\left\{
( F_{33} - F_{11} ) 
c_{\phi} s_{\phi} \left( e^{ - i h_{3} x } + e^{ - i h_{1} x } \right) 
+ \left( e^{ i \delta} F_{13} + e^{ - i \delta} F_{31} \right) 
\left( s^2_{\phi} e^{ - i h_{3} x } - c^2_{\phi} e^{ - i h_{1} x } \right)
\right\} ( - i \Delta_{21} x )
\nonumber \\
&-& 
c_{\phi} s_{\phi} 
\left\{
\sin 2\phi ( F_{33} - F_{11} )
- \cos 2\phi \left( e^{ i \delta} F_{13} + e^{ - i \delta} F_{31} \right) 
\right\}
\left( \frac{ \Delta_{21} }{ h_{3} -  h_{1} } \right) 
\left( e^{ - i h_{3} x } - e^{ - i h_{1} x } \right). 
\label{tilde-S-1st-11}
\end{eqnarray}
%
%%%%%%%%%%%% S_22 %%%%%%%%%%%%%%
\begin{eqnarray}
&& 
\tilde{S}^{(1)}_{22} 
= 
F_{22} ( - i \Delta_{21} x ) e^{ - i h_{2} x }. 
\label{tilde-S-1st-22}
\end{eqnarray} 
%
%%%%%%%%%%%% S_33 %%%%%%%%%%%%%%
\begin{eqnarray}
&& 
\tilde{S}^{(1)}_{33} 
\nonumber \\ 
&=&
F_{33} 
\left( c^2_{\phi} e^{ - i h_{3} x } + s^2_{\phi} e^{ - i h_{1} x } \right) 
( -i \Delta_{21} x )
\nonumber \\
&+& 
c_{\phi} s_{\phi} 
\left\{
- c_{\phi} s_{\phi} 
\left( e^{ - i h_{3} x } + e^{ - i h_{1} x } \right) 
( F_{33} - F_{11} ) 
+ 
\left( c^2_{\phi} e^{ - i h_{3} x } - s^2_{\phi} e^{ - i h_{1} x } \right) 
\left( e^{ i \delta} F_{13} + e^{ - i \delta} F_{31} \right) 
\right\} ( -i \Delta_{21} x ) 
\nonumber \\ 
&+&
c_{\phi} s_{\phi} 
\left\{ \sin 2\phi ( F_{33} - F_{11} ) 
- \cos 2\phi 
\left( e^{ i \delta} F_{13} + e^{ - i \delta} F_{31} \right) 
\right\}
\frac{ \Delta_{21} }{ ( h_{3} -  h_{1} ) } 
\left( e^{ - i h_{3} x } - e^{ - i h_{1} x } \right). 
\label{tilde-S-1st-33}
\end{eqnarray}
%
%%%%%%%%%%%% S_12 %%%%%%%%%%%%%%
\begin{eqnarray}
\tilde{S}^{(1)}_{12}  
&=&
\left( s^2_{\phi} F_{12} + c_{\phi} s_{\phi} e^{ - i \delta} F_{32} \right) 
\left( \frac{ \Delta_{21} }{ h_{3} -  h_{2} } \right) 
\left( e^{ - i h_{3} x } - e^{ - i h_{2} x } \right)
\nonumber \\ 
&+& 
\left( c^2_{\phi} F_{12} - c_{\phi} s_{\phi} e^{ - i \delta} F_{32} \right) 
\left( \frac{ \Delta_{21} }{ h_{2} -  h_{1} } \right) 
\left( e^{ - i h_{2} x } - e^{ - i h_{1} x } \right). 
\label{tilde-S-1st-12}
\end{eqnarray} 
%
%%%%%%%%%%%% S_21 %%%%%%%%%%%%%%
\begin{eqnarray} 
\tilde{S}^{(1)}_{21} 
&=& 
\left( s^2_{\phi} F_{21} + c_{\phi} s_{\phi} e^{ i \delta} F_{23} \right) 
\left( \frac{ \Delta_{21} }{ h_{3} -  h_{2} } \right) 
\left( e^{ - i h_{3} x } - e^{ - i h_{2} x }  \right) 
\nonumber \\ 
&+& 
\left( c^2_{\phi} F_{21} - c_{\phi} s_{\phi} e^{ i \delta} F_{23} \right) 
\left( \frac{ \Delta_{21} }{ h_{2} -  h_{1} } \right) 
\left( e^{ - i h_{2} x } - e^{ - i h_{1} x }  \right). 
\label{tilde-S-1st-21}
\end{eqnarray}
%
%%%%%%%%%%%% S_23 %%%%%%%%%%%%%%
\begin{eqnarray}
\tilde{S}^{(1)}_{23} &=& 
\left( c_{\phi} s_{\phi} e^{ - i \delta} F_{21} + c^2_{\phi} F_{23} \right) 
\frac{ \Delta_{21} }{ ( h_{3} -  h_{2} ) } 
\left( e^{ - i h_{3} x } - e^{ - i h_{2} x } \right) 
\nonumber \\ 
&-&
\left( c_{\phi} s_{\phi} e^{ - i \delta} F_{21} - s^2_{\phi} F_{23} \right) 
\frac{ \Delta_{21} }{ ( h_{2} -  h_{1} ) } 
\left( e^{ - i h_{2} x } - e^{ - i h_{1} x } \right). 
\label{tilde-S-1st-23}
\end{eqnarray}
%
%%%%%%%%%%%% S_32 %%%%%%%%%%%%%%
\begin{eqnarray}
\tilde{S}^{(1)}_{32} &=&
\left( c_{\phi} s_{\phi} e^{ i \delta} F_{12} + c^2_{\phi} F_{32} \right) 
\frac{ \Delta_{21} }{ ( h_{3} -  h_{2} ) } 
\left( e^{ - i h_{3} x } - e^{ - i h_{2} x } \right) 
\nonumber \\ 
&-& 
\left( c_{\phi} s_{\phi} e^{ i \delta} F_{12} - s^2_{\phi} F_{32} \right)
\frac{ \Delta_{21} }{ ( h_{2} -  h_{1} ) }
\left( e^{ - i h_{2} x } - e^{ - i h_{1} x } \right). 
\label{tilde-S-1st-32}
\end{eqnarray}
%
%%%%%%%%%%%% S_13 %%%%%%%%%%%%%%
\begin{eqnarray}
&& 
%% e^{ i \delta} 
\tilde{S}^{(1)}_{13} 
\nonumber \\ 
&=& 
e^{ - i \delta} 
\biggl[
F_{33} 
c_{\phi} s_{\phi} \left( e^{ - i h_{3} x } - e^{ - i h_{1} x } \right) 
( -i \Delta_{21} x ) 
\nonumber \\ 
&-&
c_{\phi} s_{\phi} 
\left\{
( F_{33} - F_{11} ) \left( s^2_{\phi} e^{ - i h_{3} x } - c^2_{\phi} e^{ - i h_{1} x } \right) 
- c_{\phi} s_{\phi} \left( e^{ i \delta} F_{13} + e^{ - i \delta} F_{31} \right)
\left( e^{ - i h_{3} x } + e^{ - i h_{1} x } \right) 
\right\}
( -i \Delta_{21} x ) 
\nonumber \\
&+& 
\left\{ 
e^{ i \delta} F_{13} 
- c_{\phi} s_{\phi} \left[ 
\cos 2\phi ( F_{33} - F_{11} ) 
+ \sin 2\phi \left( e^{ i \delta} F_{13} + e^{ - i \delta} F_{31} \right) \right] 
\right\}
\left( \frac{ \Delta_{21} }{ h_{3} -  h_{1} } \right) 
\left( e^{ - i h_{3} x } - e^{ - i h_{1} x } \right) 
\biggr].
\nonumber \\
\label{tilde-S-1st-13}
\end{eqnarray}
\begin{eqnarray}
&& 
%% e^{ - i \delta} 
\tilde{S}^{(1)}_{31} 
\nonumber \\ 
&=& 
e^{ i \delta} 
\biggl[
F_{11} 
c_{\phi} s_{\phi} \left( e^{ - i h_{3} x } - e^{ - i h_{1} x } \right) 
( -i \Delta_{21} x ) 
\nonumber \\ 
&+& 
c_{\phi} s_{\phi} 
\left\{
( F_{33} - F_{11} ) 
\left( c^2_{\phi} e^{ - i h_{3} x } - s^2_{\phi} e^{ - i h_{1} x } \right) 
+ c_{\phi} s_{\phi} 
\left( e^{ i \delta} F_{13} + e^{ - i \delta} F_{31} \right) 
\left( e^{ - i h_{3} x } + e^{ - i h_{1} x } \right) 
\right\} 
( -i \Delta_{21} x ) 
\nonumber \\
&+& 
\left\{
e^{ - i \delta} F_{31} 
- c_{\phi} s_{\phi} \left[ \cos 2\phi ( F_{33} - F_{11} ) 
+ \sin 2\phi \left( e^{ i \delta} F_{13} + e^{ - i \delta} F_{31} \right) 
\right] 
\right\} 
\left( \frac{ \Delta_{21}  }{ h_{3} -  h_{1} } \right) \left( e^{ - i h_{3} x } - e^{ - i h_{1} x } \right) 
\biggr]. 
\nonumber \\
\label{tilde-S-1st-31}
\end{eqnarray}

Then, by using $S = U_{23} \tilde{S} U_{23}^{\dagger}$ in eq.~\eqref{hatS-S-relation}, and for the given $\tilde{S}$ matrix elements, the flavor basis $S$ matrix elements can be written as 
\begin{eqnarray} 
S_{ee} &=& \tilde{S}_{11}, 
\nonumber \\
S_{e \mu} &=& c_{23} \tilde{S}_{12} + s_{23} \tilde{S}_{13}, 
\nonumber \\ 
S_{e \tau} &=& c_{23} \tilde{S}_{13} - s_{23} \tilde{S}_{12},
\nonumber \\
S_{\mu e} &=& c_{23} \tilde{S}_{21} + s_{23} \tilde{S}_{31} 
= S_{e \mu} (- \delta),  
\nonumber \\
S_{\mu \mu} &=& c^2_{23} \tilde{S}_{22} + s^2_{23} \tilde{S}_{33} + c_{23} s_{23} ( \tilde{S}_{23} + \tilde{S}_{32} ), 
\nonumber \\
S_{\mu \tau} &=& c^2_{23} \tilde{S}_{23} - s^2_{23} \tilde{S}_{32} + c_{23} s_{23} ( \tilde{S}_{33} - \tilde{S}_{22} ), 
\nonumber \\ 
S_{\tau e} &=& c_{23} \tilde{S}_{31} - s_{23} \tilde{S}_{21} 
= S_{e \tau} (- \delta), 
\nonumber \\ 
S_{\tau \mu} &=& c^2_{23} \tilde{S}_{32} - s^2_{23} \tilde{S}_{23} + c_{23} s_{23} ( \tilde{S}_{33} - \tilde{S}_{22} ) = S_{\mu \tau} (- \delta),
\nonumber \\
S_{\tau \tau} &=& s^2_{23} \tilde{S}_{22} + c^2_{23} \tilde{S}_{33} - c_{23} s_{23} ( \tilde{S}_{23} + \tilde{S}_{32} ).
\label{tildeS-S-relation}
\end{eqnarray}

\section{The oscillation probability in the $\nu_{\mu} \rightarrow \nu_{\tau}$ channel}
\label{sec:P-mutau-matter}

Here, we present the explicit expression of $P(\nu_{\mu} \rightarrow \nu_{\tau}) = P(\nu_{\mu} \rightarrow \nu_{\tau})^{ \text {non-int-fer} } + P(\nu_{\mu} \rightarrow \nu_{\tau})^{ \text {int-fer} }$, in which the $\phi$ symmetry is manifest.
\begin{eqnarray} 
&&
P(\nu_{\mu} \rightarrow \nu_{\tau})
\nonumber \\ 
&=& 
4 c^2_{23} s^2_{23} \left[
- c^2_{\phi} s^2_{\phi} 
\sin^2 \frac{ ( h_{3} - h_{1} ) x }{2} 
+ c^2_{\phi} 
\sin^2 \frac{ ( h_{3} - h_{2} ) x }{2} 
+ s^2_{\phi} 
\sin^2 \frac{ ( h_{2} - h_{1} ) x }{2} 
\right]
\nonumber \\ 
&+& 
2 c^2_{23} s^2_{23} 
\biggl[ 
c^2_{\phi} s^2_{\phi} s^2_{12} \cos 2 ( \phi - \theta_{13} )
\sin ( h_{3} - h_{1} ) x 
\nonumber \\
&-& 
c^2_{\phi} \left[ c^2_{12} - s^2_{12} \sin^2 ( \phi - \theta_{13} ) 
\right] 
\sin ( h_{3} - h_{2} ) x 
+ s^2_{\phi} \left[ c^2_{12} - s^2_{12} \cos^2 ( \phi - \theta_{13} ) \right]
\sin ( h_{2} - h_{1} ) x 
\biggr] ( \Delta_{21} x ) 
\nonumber \\ 
&-& 
4 c^2_{23} s^2_{23} s^2_{12} c_{\phi} s_{\phi} 
\sin 2 ( \phi - \theta_{13} ) 
\frac{ \Delta_{21} }{ ( h_{3} -  h_{1} ) } 
\nonumber \\ 
&\times& 
\biggl[
\cos 2\phi \sin^2 \frac{ ( h_{3} - h_{1} ) x }{2} 
+ \sin^2 \frac{ ( h_{3} - h_{2} ) x }{2} 
- \sin^2 \frac{ ( h_{2} - h_{1} ) x }{2} 
\biggr]
\nonumber \\ 
&+& 
4 \tilde{J}_{mrs}^{c} \cos 2\theta_{23} \cos \delta
\frac{ \Delta_{21} }{ ( h_{2} -  h_{1} ) }
%\nonumber \\ &\times&
\left\{
c^2_{\phi} \sin^2 \frac{ ( h_{3} - h_{2} ) x }{2} 
+ ( 1 + s^2_{\phi} ) \sin^2 \frac{ ( h_{2} - h_{1} ) x }{2} 
- c^2_{\phi} \sin^2 \frac{ ( h_{3} - h_{1} ) x }{2} 
\right\} 
\nonumber \\
&+& 
4 \tilde{J}_{mrs}^{s} \cos 2\theta_{23} \cos \delta
\frac{ \Delta_{21} }{ ( h_{3} -  h_{2} ) }
%\nonumber \\ &\times&
\left\{
( 1 + c^2_{\phi} ) 
\sin^2 \frac{ ( h_{3} - h_{2} ) x }{2} 
+ s^2_{\phi} \sin^2 \frac{ ( h_{2} - h_{1} ) x }{2} 
- s^2_{\phi} \sin^2 \frac{ ( h_{3} - h_{1} ) x }{2} 
\right\}
\nonumber \\
&+&
8 \sin \delta 
\left[ 
\tilde{J}_{mr}^{c} \frac{ \Delta_{21} }{ ( h_{2} -  h_{1} ) }
- \tilde{J}_{mr}^{s} \frac{ \Delta_{21} }{ ( h_{3} -  h_{2} ) } 
\right] 
%\nonumber \\ &\times&
\sin \frac{ ( h_{3} - h_{1} ) x }{2} 
\sin \frac{ ( h_{2} - h_{1} ) x }{2} 
\sin \frac{ ( h_{3} - h_{2} ) x }{2}, 
\label{helio-P-total-P}
\end{eqnarray}
where $\tilde{J}_{mr}^{s} $ etc. are defined in eqs.~\eqref{matter-Jarlskog-2nd} and \eqref{matter-Jarlskog-3rd}. We note that the following identities are useful.
\begin{eqnarray} 
&& 
\left[ \sin ( h_{3} - h_{2} ) x - \sin ( h_{3} - h_{1} ) x 
+ \sin ( h_{2} - h_{1} ) x \right] 
= 4 \sin \frac{ ( h_{3} - h_{1} ) x }{2} \sin \frac{ ( h_{2} - h_{1} ) x }{2} 
\sin \frac{ ( h_{3} - h_{2} ) x }{2},
\nonumber \\ 
&&
- \sin^2 \frac{ ( h_{3} - h_{2} ) x }{2} 
+ \sin^2 \frac{ ( h_{3} - h_{1} ) x }{2} 
+ \sin^2 \frac{ ( h_{2} - h_{1} ) x }{2} 
= 2 \sin \frac{ ( h_{2} - h_{1} ) x }{2} 
\sin \frac{ ( h_{3} - h_{1} ) x }{2}
\cos \frac{ ( h_{3} - h_{2} ) x }{2}.
\nonumber \\
\label{identities}
\end{eqnarray}

\end{document}